\documentclass[fleqn,usenatbib]{mnras}
\usepackage[normalem]{ulem}
\usepackage{newtxtext}
\usepackage[varvw]{newtxmath}

\usepackage[T1]{fontenc}
\DeclareRobustCommand{\VAN}[3]{#2}
\let\VANthebibliography\thebibliography
\def\thebibliography{\DeclareRobustCommand{\VAN}[3]{##3}\VANthebibliography}

\usepackage[utf8]{inputenc}
\usepackage[normalem]{ulem}
\usepackage[caption = false]{subfig}
\usepackage{lineno}

\usepackage{graphicx}	
\usepackage{amsmath}	

\title[Cosmic ray diffusion subgrid models]{Comparing subgrid models for cosmic ray diffusion in a magnetized isolated galaxy simulation}

\author[Sarah Thiele]{
Sarah Thiele$^{1}$\thanks{E-mail: sarah.thiele@princeton.edu}
and Romain Teyssier$^{1}$
\\
$^{1}$Department of Astrophysical Sciences, Princeton University, 4 Ivy Lane, Princeton, NJ 08544, USA \\
}

\date{Accepted XXX. Received YYY; in original form ZZZ}

\pubyear{2015}

\begin{document}
\label{firstpage}
\pagerange{\pageref{firstpage}--\pageref{lastpage}}
\maketitle

\begin{abstract}
Galactic cosmic rays (CRs) play a crucial role in galaxy formation and evolution by altering gas dynamics and chemistry across multiple scales. Typical numerical simulations of CR transport assume a constant diffusion coefficient for the entire galaxy, despite both numerical and theoretical studies showing that it can change by orders of magnitude depending on the phase of the interstellar medium. Only a few simulations exist that self-consistently calculate CR transport with diffusion, streaming, and advection by the background gas. In this study we explore three subgrid models for CR diffusion, based on popular theories of CR transport. We post-process an isolated, star-forming MHD galactic disk simulated using the \texttt{RAMSES} code. The resulting diffusion coefficients depend solely on the subgrid turbulent kinetic energy and the MHD state variables  of the plasma. We use these models to calculate coefficients for vertical transport. We find that they depend critically on the local magnetic field tilt angle. Across models, our resulting diffusion coefficients range from $10^{26}~\rm cm^2s^{-1}$ to $10^{31}~\rm cm^2s^{-1}$, and yield CR energy densities at the midplane from $1$ to $100 ~\rm eV cm^{-3}$, suggesting varied degrees of backreaction on their environment. Using simple approximations, we show that the gamma ray luminosity of the galaxy depends primarily on the gas surface density and the turbulent confinement of CRs by the galactic corona.
\end{abstract}




\section{Introduction}\label{sec:intro}
Cosmic Rays (CRs) are non-thermal particles that are produced in high-energy astrophysical events like supernovae, pulsar winds and active galactic nuclei.  In particular, the CR energy spectrum is dominated by particles that originate within the Milky Way (``galactic cosmic rays") and have energies around 1~GeV per nucleon. At these energies, the relative abundances are dominated by CR protons and alpha particles, theorized to be produced by core-collapse supernovae \citep{Friedlander, Drury2012, Marcowith2018}, where the CRs are accelerated via Fermi acceleration.

CRs are also an important component of the Interstellar Medium (ISM): measurements of their energy densities have revealed values around $1~\rm eV cm^{-3}$, in approximate equipartition with turbulent, thermal, magnetic and radiative energy \citep{Boulares}. Consequently, CR pressure gradients may alter gas dynamics in the ISM across a multitude of scales \citep{grenier2015, commercon19}. On parsec scales, CRs determine the thermal energy and ionisation state of the dense ISM. CRs are the dominant heating and ionizing mechanism in the cores of molecular clouds since these clouds are often shielded against many photons due to dust. They induce spallation reactions that create unstable secondary CRs and constrain the residency time of CRs in the Galaxy around 15-20 Myr \citep{Ceccarelli, commercon19, PhysRevD.101.023013}. 

On kiloparsec scales, CR pressure provides vertical support against gravity in addition to thermal pressure from hot gas. This could thicken the galactic disc \citep{Girichidis2016, NC2022}, and help support cool gas in the CGM, increasing its volume-filling fraction instead of it remaining confined to dense filaments \citep{ji2020}. CR pressure can also help to resist gas inflow and regulate the gas accretion rate onto the Galaxy \citep{Butsky2020}.  

CRs can also drive galactic winds out of Milky Way-like galaxies and possibly strip entire dwarf galaxies. The combination of suppressing gas accretion and driving gas outflows could significantly affect star formation in the Galaxy by regulating baryon cycling between diffuse gas reservoirs like the warm and hot ISM and the circumgalactic medium (CGM), and dense star-forming clouds (see review by \citealt{Recchia2020}. Also see \citealt{booth2013, Hanasz2013, salem2014, ruszkowski2017, DD2020, Hopkins2021, NC2022, ostriker24}). This is why CRs have been identified as a potential crucial player in galaxy evolution and star formation. However, they are still a topic of ongoing research with many questions still unanswered, primarily focused on our poor understanding on how CRs are transported on different scales in the galaxy.

\subsection{Mechanisms for cosmic ray transport}
\label{sec:mechanism}

Here we describe two different mechanisms for CR transport. It is known that the propagation and acceleration of CRs are governed by their interactions with magnetic fields and how they are coupled to the background thermal gas. Because they are charged, CRs will ``stream" along magnetic field lines by gyrating under the influence of the Lorentz force. However, astrophysical magnetic fields are also turbulent, and CRs will undergo resonant and non-resonant interactions with magnetic fluctuations on scales comparable to the CR's Larmor radius. Specifically, interactions of CRs with MHD turbulence is believed to be the principal mechanism for CR pitch-angle scattering, which causes their momentum distribution to become isotropic via a random walk perpendicular to the magnetic field lines \citep{Jokipii1966} (even though CRs are probably injected into the ISM anisotropically).

There is uncertainty about whether these magnetic fluctuations are Alfvén waves generated by the CRs themselves through the streaming instability (for which CRs are deemed ``self-confined"), or through external fluctuations caused by the existing background turbulence. There is evidence that the primary origin of the magnetic fluctuations may depend on the energy of the CR; specifically, it has been found that $\sim$ GeV CRs are predominantly scattered by self-excited waves \citep{blasi2012, zweibel2013, zweibel2017, evoli2018, ostriker22}. If one is to describe CR diffusion as an effective macroscopic process, integrating over these microscopic scattering processes, a correct formulation of CR transport in the vicinity of their origin sites (supernovae, pulsar winds, etc.) would depend on correctly describing the properties of the local turbulence in these regions.

Turbulence in star-forming regions of the galactic disk is highly supersonic and super-Alfvénic, with a turbulent injection scale, which corresponds to the largest eddies in the system, reaching up to $100~\rm pc$ \cite{breit2017}. At these scales (much larger than the typical CR gyro-radius for the GeV CRs that dominate the CR energy budget), it has been found that CR transport can also be modelled as a diffusive process driven extrinsically by interstellar turbulence. As described in \cite{Sam22}, the turbulent nature of the ISM means that the geometry of magnetic field lines can by highly curved and time-dependent. \cite{Sam22} claimed that while there may be random walk-like processes affecting CR transport perpendicular to the local field lines on small scales (as described above), the effective transport of CRs on scales of a pc and above can be effectively modelled as a diffusive process depending on the local properties of the medium.

Scattering that occurs in the self-confinement scenario can cause the CR streaming speed to be restricted to the ion Alfvén speed, although wave damping can suppress this effect so that CR transport can exceed the ion Alfvén speed \citep{Skilling1971, Wiener_2013} and therefore the transport speed becomes energy dependent \citep{Holmes1974}. 

A complete picture of CRs propagation involves a combination of streaming along the magnetic field, diffusion relative to Alfvén waves, and advection by background gas \citep[see][for further discussion on comparing CR transport theory with observations]{Kempski22}. Putting this all together, one can model CR transport on large scales without resolving the microscopic pitch angle scattering processes by instead just assuming that self-confinement applies such that CRs are streaming at the ion Alfvén speed, and combine this with transport caused by large-scale turbulence and gas advection to simulate effectively diffusive transport.

As CR transport has been shown to depend on local plasma conditions, it is therefore also dependent on the phase of the ISM. The warm and hot ionized ISM make up the highest volume-filling fraction in the galaxy, whereas most of the mass lies in the cold and warm neutral mediums. Depending on the ISM phase, gas turbulence might dominate CR transport over streaming and self-confinement, or vice versa. For example, wave damping suppresses the streaming instability for areas with a low ionization fraction \citep{ostriker22, Kempski22}. This will be discussed further in Section \ref{subsec:phase}. 

\subsection{Cosmic ray diffusion in galaxy simulations}

The evolution of the CR energy density in a galaxy depends on how efficiently CRs are transported. This is traditionally modelled using a single (isotropic) diffusion coefficient $\kappa$ (in units of $\rm cm^2~s^{-1}$) for the entire galaxy. In most models and simulations, the diffusion coefficient is left as a free parameter. This parameter is usually tuned to reproduce the gamma ray emission of said galaxy \cite{Hopkins_2023}. Indeed, when CR protons collide with thermal protons in the interstellar medium, they produce neutral and charged pions, the former of which then decay into pairs of gamma rays. This gamma-ray emission (usually in the GeV-TeV range) is therefore used as a probe for the underlying galactic CR energy density \citep{refId0}.

Simulations of galactic CR transport vary in the scale of the system they simulate; how CR transport is incorporated into the simulation; and the diffusion coefficient model used. For reviews of CR transport mechanisms and simulations, see \citet{Amato_2018} and \citet{3}.

Studies range in simulation scale from a patch of the ISM \citep[e.g.][which simulate and analyze box sizes of $2\times2\times40~\rm{kpc}$ and $1\times1\times7~\rm{kpc}$ respectively]{Farber_2018, armillotta21}, to isolated LMC/dwarf-sized galaxies \citep{Girichidis_2021, DD2020, MartinAlvarez2023}, to isolated $L_*$ or starburst galaxies \citep[][]{salem2014, ruszkowski2017, ji2020, Chiu_2024}. In addition, cosmological zoom-ins have studied how CRs affect galaxies at different masses and redshifts \citep[see e.g.][]{Hopkins_2020, Montero2024}. The results of simulation-based studies can vary. For example, the cosmological zoom-ins of \citet{Hopkins_2020} found that CRs do not have a significant impact on galaxy properties in dwarf galaxies or at high redshifts, but can suppress star formation at higher galaxy masses and lower redshifts. Conversely, the isolated galaxy simulations of \citet{Jacob2017} found that mass-loaded winds were only driven beyond the virial radius by CRs in lower mass ($< 10^{12}M_\odot$) haloes.

CR transport is often incorporated using a two-fluid approximation in mesh-based hydrodynamic/MHD codes, where CRs are modelled as an ultra-relativistic fluid with adiabatic index $\gamma_{\rm CR}=4/3$ and the thermal gas is given $\gamma_{\rm gas}=5/3$ \citep[e.g.][]{hanasz2003, Farber_2018}. This might be modeled in a post-processing scheme \citep[e.g.][]{ostriker22}, or with the CR energy density simulated self-consistently to study CR backreaction \citep[e.g.][]{booth2013, salem2014, Chan_2019, thomas2023, ostriker24}. Furthermore on CR feedback, \citet{jiang2018} used a two-moment method to model streaming CRs that improved upon numerical stability. \citet{thomas2019} derived equations for CR transport, also with a two-moment method, that improved upon the treatment of CR scattering; these were implemented into a finite volume scheme for CR hydrodynamics in \citet{thomas2021}. \citet{Hopkins_2023} proposed a simplified subgrid model for CR feedback, wherein the diffusion coefficient is tuned to reproduce CR observables. This sub-grid model was implemented in \citet{Ramesh_2024}. They found that depending on the galaxy mass, CRs suppress galaxy stellar masses for a fixed halo mass. Furthermore, they found that CRs affected the pressure structure of the halo gas more significantly than gas in the CGM.

Simulation results depend on which CR transport mechanisms out of streaming, advection, and diffusion are taken into account (e.g. \citealt{booth2013} and \citealt{Hanasz2013} model advection + diffusion, whereas \citealt{Chan_2019} compares different combinations of transport mechanisms). It is widely accepted that correctly dynamically coupling the CR population to the underlying thermal plasma is important for self-consistently simulating CR transport, and accounting for all CR transport mechanisms is necessary to correctly model all CR effects \citep{Butsky_2018, 3}. For example, \citet{Uhlig_2012} modeled streaming and advection and found that advection alone is not sufficient to induce galactic winds.

Some simulations of a patch of local ISM have done physically-motivated calculations of diffusion coefficients - for example, \citet{armillotta21} calculated the CR scattering coefficient by balancing wave growth and damping and applied it in a post-processing scheme. However, as noted above, most models assume a diffusion coefficient that is constant throughout the galaxy \citep[e.g.][]{DC2016, commercon19, Rodríguez2021, NC2022}. Either isotropic or anisotropic diffusion is assumed, and the diffusion coefficient is typically chosen between $10^{27}$ and $10^{31}~\rm cm^2 s^{-1}$. \citet{Girichidis_2021} and \citet{1b} instead apply a spectral model with an energy-dependent diffusion coefficient and find that effective diffusion coefficients in spectral models vary by multiple orders of magnitude.

Across studies involving CR transport in galaxy simulations, there are consistent takeaways. Firstly, CRs suppress star formation and thus decrease the stellar mass of a galaxy. This is due to the CRs providing pressure support for cold, low-density gas to keep it from collapsing into dense filaments where stars are formed. They furthermore accelerate galactic outflows, and these outflows are cooler and denser than those driven by other mechanisms. Lastly, most studies suggest that just accounting for one type of CR transport is insufficient to capture all CR effects, and similarly for not taking into account ISM phase (e.g. CRs can decouple from the ISM in the cold neutral medium phase which leads to different outflow behaviour).

Numerical solvers for the streaming instability and CR acceleration has been implemented using a fluid approach in RAMSES, where both streaming and anisotropic diffusion are taken into account \citep[e.g.][]{Dubois2019, DD2020, NC2022, MartinAlvarez2023, Montero2024}, however these models all assume constant diffusion coefficients across the galaxy.

In order to correctly model the effect of CRs on galaxy evolution, it is crucial to define the diffusion coefficient correctly, and calculate it self-consistently based on the local physical properties of the plasma. Furthermore, a self-consistent simulation reflects the impact of the ISM's multi-phase nature on CR transport that a single diffusion coefficient neglects \citep{ostriker22, ostriker24}. In this work, we describe a new methodology to calculate CR diffusion coefficients that can be implemented as two different subgrid models for CR diffusion; these diffusion coefficients take into account the local variations of the gas in different regions of a galaxy.

Our first model is based on an analytic formulation for CR transport motivated by averaging over pitch-angle scattering processes, hereafter called our ``microscopic model''. Although in this work we make many simplifying assumptions over previous calculations, we believe our model yields realistic values for the diffusion coefficients, quite comparable to those found in the literature \citep[see for examples e.g.][or review by \cite{LazReview}]{Chandran2000, YL02, YL04, YL08}.

Our second model calculates CR diffusion coefficients where diffusive motion is driven by self-confined, streaming CRs transported by large-scale turbulence, hereafter our ``macroscopic model". For this model we employ empirical fits by \citet{Sam22}, hereafter S23, which utilizes MHD turbulence simulations to study large-scale CR transport. 

Both approaches yield diffusion coefficients that are functions dependent solely on properties of the plasma in the galactic disc like gas density $\rho$, ionization fraction $\chi$ and Alfvén Mach number $\mathcal{M}_{A}$, which makes it easy to implement in any MHD simulation. As a proof of concept, we apply our subgrid models to an isolated star-forming galaxy simulated with the \texttt{RAMSES} code \citep{ramses} as a post-processing scheme.

We describe in more detail each model in Section~\ref{sec:modelmicro} and Section~\ref{sec:modelmacro}. We define a critical diffusion coefficient set by advection of CRs via hydrodynamical turbulence in Section~\ref{sec:critical}. This critical value limits the regimes where CRs remain trapped by turbulence within a particular length scale. Any effective model of CR transport is based on some kind of averaging procedure over smaller, unresolved scales; we describe different averaging and weighting schemes in Section~\ref{sec:avg}. In Section~\ref{sec:diffcoeff}, we analyze the resulting diffusion coefficients, parallel and perpendicular to the local magnetic field, in each model and discuss the implications that these values have on the overall transport of CR energy density within the galaxy. 

Recent studies \citep[e.g.][]{Sam22} have shown that CR transport may be super-diffusive instead of diffusive. Super-diffusion arises when the underlying random walk cannot be describe using fixed coefficients. We model super-diffusion using a scale-dependent effective diffusion coefficient and discuss its effect on CR transport in Section~\ref{sec:superdiff}. In Section~\ref{sec:kappaz}, we compute the vertical component of the diffusion coefficient (pointing out of the plane of the disc), $\kappa_z$, to quantify CR transport out of the galaxy. We discuss the dependence of the diffusion coefficient on the ISM gas phase in Section~\ref{subsec:phase}. Finally, we use our $\kappa_z$ calculations to estimate the gamma-ray emission and the midplane CR energy density associated with each model in Section~\ref{sec:cal} and conclude in Section~\ref{sec:conclusions}. 

Note that our post-processing approach does not model the backreaction of CR transport to the evolution of the galaxy. We intend to compare our two subgrid models using a fully self-consistent MHD simulation coupled to CR transport using the \texttt{RAMSES} code in a follow-up paper.

\section{Methods}\label{sec:methods}

In this paper, we analyze a simulation performed using the \texttt{RAMSES} code and presented in a previous paper by \cite{Girma}. As described therein, this simulation features an isolated Milky Way-like galaxy with initial conditions coming from the \textit{AGORA} project \citep{kim2016}, a project that compares high-resolution simulations of galaxy formation across different codes. 

Our galaxy includes initially stars in the galactic disk and in the bulge, gas in the disk and in the surrounding halo, and a dark matter halo. The gaseous component is modeled using the ideal MHD equations with a sub-grid model for turbulence based on the Large Eddy Simulation (LES) framework. As described in Section~3 of \cite{Girma}, the disk gas has been initialized with a resolved turbulence flow by adding to the rotation of the disk random velocity fluctuations that follow Burgers turbulence (more on Burger's turbulence in the following section). Note that this initial turbulence quickly decays and resolved turbulence in the disk is maintained via stellar feedback and disk instabilities.

As described in Section~2 of \cite{Girma}  \citep[also see references within including][]{Schmidt_2014, Semenov2016, Kretschmer2019}, \texttt{RAMSES} tracks the evolution of the the sub-grid turbulent kinetic energy, from which one can solve for the one-dimensional turbulence velocity dispersion $\sigma_{\rm{1D}}$ at the driving scale of turbulence (i.e. the cell size $L_{\rm{min}}$), encapsulating turbulent motion below the grid scale. This LES model is key to our study because having information about the MHD properties of the local plasma and about turbulence on unresolved scales is precisely what allows us to derive a sub-grid model for CR transport.

The galaxy's magnetic field has been initialized with a constant magnetic field of $10^{-20}~\rm{G}$ pointing in the positive-$z$ direction. This very weak magnetic field then grows exponentially owing to the combination of a subgrid dynamo model and the resolved turbulence and rotation of the galaxy \citep{Liu.2022}.

The galaxy's dark matter halo has a mass of $10^{12}~M_\odot$ and the disk has a mass of $4\times 10^{10}~M_\odot$ with gas fraction $f_{\rm gas} = 0.2$. The disk has an exponential density profile in both the radial and vertical directions with a disk scale length $R_d = 3.4~\rm{kpc}$ and scale height $H_d = 0.1 R_d$. The adaptive mesh refinement of \texttt{RAMSES} is set to have a minimum cell size (i.e. the finest resolution size) of $L_{\rm{min}}= 20 \rm{pc}$ \citep[more details are given in][]{Girma}.

\subsection{Microscopic Model: a simple analytic formulation for cosmic ray diffusion}
\label{sec:modelmicro}

As discussed in Section \ref{sec:mechanism}, our microscopic model is based on an effective theory designed by averaging the microscopic effect of pitch-angle scattering. CRs propagating away from their injection site perform a random walk in momentum space as they scatter off of random magnetic fluctuations, leading to diffusive transport. As previously noted, these fluctuations may be Alfvén waves resulting from the streaming instability, or fluctuations in the background plasma from other external processes. 

While motion parallel to the magnetic field arises as a natural consequence of CRs being charged particles and thus gyrating along field lines, transport perpendicular to the mean magnetic field is dominated by this random walk. Alternatively if perpendicular transport is sub-diffusive instead (from CRs often scattering backwards along their trajectories), perpendicular transport of the field lines themselves through advection could be an alternative mechanism of producing diffusion perpendicular to the mean magnetic field \citep{YL08}.)

Traditional calculations of CR diffusion coefficients use quasi-linear theory (QLT), where CR--wave interactions are treated with perturbation theory \cite[see][and references therein]{LazReview}. We model our analytic formulation for diffusion after \cite{YL08} (hereafter YL08). YL08 extends QLT and performs calculations of the CR mean free paths (MFP) in order to determine CR propagation both parallel and perpendicular to the mean magnetic field. 

In this section, we describe a simplified version of YL08 theory that we believe captures most of the quantitative predictions. We begin by assuming that the CRs are moving at the speed of light $c$ in the fluid frame of reference, gyrating around a magnetic field line. YL08 separates the diffusion coefficient calculations into high and low Alfvén Mach number regimes separated by $\mathcal{M}_{A}=1$, which we do as well. 

Next we examine the different scale lengths involved. YL08 performed detailed calculations of the MFP $\lambda_\parallel$ parallel to the mean magnetic field. These calculations require a complex integration of scattering coefficient prescriptions over wavenumber and pitch angle. 

However, the MFP only enters the final diffusion coefficient calculations in the case that the MFP is below a critical length scale $L$, where for theoretical models typically $L=L_{\rm{inj}}$ with $L_{\rm{inj}}$ the injection (or ``driving'') scale of the turbulence. 

We assume that for our CRs of interest with $\sim$ GeV energies, $\lambda_\parallel \geq L$ over the full $\mathcal{M}_{A}$ regime. Indeed, previous studies have found that for instance in the warm and cold neutral mediums (WNM/CNM), GeV CRs have Larmor radii below MHD turbulence damping scales and so therefore are in the regime $\lambda_\parallel \gg L_{\rm{inj}}$ \citep{Xu16, commercon19}. For model YL08, we take $L_{\rm{inj}}=L_{\rm{min}}$, where $L_{\rm{min}}$ is the minimum cell size in the \texttt{RAMSES} grid at a given time step dictated by the adaptive mesh refinement (AMR) strategy. We argue that this resolution dependence should not impact our results, provided that the simulation resolves at least the beginning of the turbulent cascade; an increase in simulation resolution corresponds to a $\Delta x^{1/2}$ decrease in subgrid turbulence ($\Delta x$ being the grid size), such that the the model will deliver the same result.

We use the notation of $\kappa_\parallel$ and $\kappa_\perp$ for the diffusion coefficient parallel and perpendicular to the local magnetic field, with superscripts ``micro" and ``macro" to denote which model we are referring to in plots where both models are shown. 

In the super-Alfvénic regime ($\mathcal{M}_A > 1$),  CRs are traveling at speed $v\sim c$, the speed of light, and see themselves traveling in a straight line on length scales below the Alfvén scale $\ell_A$ in all directions. The timescale for pitch angle scatterings $\tau$ in the parallel direction has associated frequency $\tau^{-1}=\nu=\Omega\Big{(}\frac{\delta B_0}{B_0}\Big{)}^2$, where $B_0=|\vec{B}|$ is the magnitude of the background magnetic field, and $\Omega= c/2\pi R$ is the gyrofrequency with $R$ the gyroradius.  This leads to the parallel diffusion coefficient \begin{equation}
    \kappa_\parallel = \frac{c^2}{\nu} = \frac{c^2}{\Omega}\Bigg{(}\frac{\delta B_0}{B_0}\Bigg{)}^{-2}
\end{equation} 

At wavenumber scales $\ell$ below some $\ell_0^{-1}$, fluctuations in the magnetic field $\delta B_0$ are given by $\delta B_0^2(\ell)/8\pi= \frac{1}{2} \rho \delta v^2(\ell)$ if we assume that the magnetic field fluctuations are in equipartition with the velocity field fluctuations $\delta v$. $\delta v$ depends on the length scale and velocity dispersion of the fluid with $\delta v^2=3\sigma_{\rm 1D}^2(\ell/\ell_0)$, using Burger's turbulence and taking into account three dimensions. We next note that the Alfvén speed is $v_A =  \frac{B_0}{\sqrt{4\pi\rho}}$ and Alfvén Mach number is $\mathcal{M}_{A}= \frac{\sigma_{\rm 1D}\sqrt{4\pi\rho}}{B_0}$. Putting in everything for $\Omega$, $\delta B_0^2$ and $B_0^2$ and rearranging we have \begin{align}
    \kappa_\parallel &= \frac{2\pi Rc\ell_0}{3 \mathcal{M}_A^2\ell}
\end{align} and because our length scales of interest for this model are comparable to the gyroradius, we set $\ell = 2\pi R$. In our case $\ell_0=L_{\rm inj}$. We can write $\kappa_\parallel$ in terms of the Alfvèn scale $\ell_A=L_{\rm{inj}}/\mathcal{M}_A^2$, the scale at which the turbulence becomes trans-Alfvénic. Lastly, we note that in the super-Alfvénic regime, turbulence isotropises motion in the direction perpendicular to the magnetic field \citep{critbal}, so the diffusion is isotropic and we have $\kappa_\perp=\kappa_\parallel$. Altogether, the equation we use is

\begin{equation}\label{eq:microiso}
\kappa_\parallel=\kappa_\perp=\frac{\ell_A c}{3}, \ \ \ \mathcal{M}_{A} > 1
\end{equation}

Note  that we modify the $\ell_A$ scale that is used in e.g. YL08 and \cite{commercon19} (which have $\ell_A\propto \mathcal{M}_A^{-3}$ following Kolmogorov turbulence) according to Burgers turbulence, which yields a $\mathcal{M}_A^{-2}$ proportionality. On the large (molecular cloud to galactic-sized) scales that we are studying CR transport on, the turbulent cascade's inertial range will follow Burger's turbulence. In molecular clouds, there is a transition point where the turbulence moves from supersonic to subsonic, called the sonic scale, that induces a break in the turbulence power spectrum from Burger's $E(k)\propto k^{-2}$ to Kolmogorov $E(k)\propto k^{-5/3}$ \citep{fed}. We are in the supersonic regime, and the high wave number regime of the inertial range that would follow Kolmogorov turbulence operates on scales too small for us to take into consideration in this study.

Further on this point, \cite{Kempski22} includes a discussion about uncertainties in the physics behind MHD fast-mode turbulence (see their Section 5), which affect our microscopic model. One is that YL08 assumes that the fast-mode power spectrum is isotropic in the absence of wave damping, which is likely not true for low values of plasma $\beta$. Secondly, YL08 assumes that MHD fast-modes follow a weak-turbulence cascade, however it has been argued that this can be suppressed by steepening of sound waves and a subsequent formation of weak shocks \citep{zakharov1970, kadomtsev1973, Kempski22}. In this case fast modes would follow the aforementioned spectrum $\propto k^{-2}$; this has been observed by e.g. \citet{Kowal2010} and \citet{Makwana2020}. Many studies have tried to simulate CR transport including both self-confinement and turbulence \citep[e.g.][]{Aloisio2013, Aloisio2015}, but these models assume an undamped Kolmogorov cascade. See \citet{hopkins20202} for comparisons of physical models.

In the sub-Alfvénic regime, CR propagation is no longer isotropic, and in general more aligned with the local magnetic field at the scale of $L_{\rm{inj}}$ since in this case magnetic fields have a greater influence on the gas dynamics than the turbulence. The CR motion is dominated by magnetic field line motion and does not experience as much advection. We therefore assume that the parallel diffusion coefficient dominates, $\kappa_\parallel \gg \kappa_\perp$, and that $\lambda_\parallel \gtrsim L_{\rm{inj}} > \lambda_{\rm{crit}}$, where $\lambda_{\rm{crit}}$ is a critical MFP below which CRs become trapped and imply dynamical effects on the surrounding gas flow. The diffusion coefficients in this regime, as found by YL08, are given by
\begin{align}\label{eq:micropar}
\kappa_\parallel &= \frac{\ell_A c}{3}  
= \frac{L_{\rm{inj}} c}{3\mathcal{M}_A^2}, \ \ \mathcal{M}_A < 1
\end{align}
and
\begin{equation}\label{eq:microperp}
\kappa_\perp = \kappa_\parallel \mathcal{M}_A^4
= \frac{L_{\rm{inj}} c \mathcal{M}_A^2}{3}, \ \ \mathcal{M}_A < 1.
\end{equation}
This completes our simplified derivation of the diffusion coefficients in the microscopic model.

\subsection{Macroscopic Model: cosmic ray streaming via turbulent diffusion}\label{sec:modelmacro}

\begin{table*}
    \centering
    \begin{tabular}{l|c|c}
      Phase   &  Temperature (K) & Ionization fraction \\ \hline
       cold neutral medium (CNM) & $T<6\times 10^3$ & $\chi < 0.5$ \\
       warm neutral medium (WNM) & $6\times 10^3<T<3.5\cdot 10^4$ & $\chi < 0.5$ \\
       warm ionized medium (WIM) & $6\times 10^3<T<3.5\cdot 10^4$ & $\chi \geq 0.5$ \\
       warm-hot ionized medium (WHIM) & $3.5\times 10^4\leq T<5\times 10^5$ & $\chi \geq 0.5$ \\
       hot ionized medium (HIM) & $T\geq5\times 10^5$ & $\chi \geq 0.5$
    \end{tabular}
    \caption{Phases of the ISM/CGM, generalized from Table 3 of \citep{kim2023}.}
    \label{tab:phases}
\end{table*}

As introduced in Section \ref{sec:mechanism}, CRs stream along magnetic field lines in the ISM as they gyrate around them due to the Lorentz force. If the streaming speed exceeds the ion Alfvén speed $v_{Ai}$ (which is the typical speed of magnetized waves in the background plasma), the streaming instability is triggered, exciting Alfvén waves in the plasma \citep{PARKER1964735, Skilling1975, Amato2011, Dubois2019, Marcowith2021, Xu2022}. Throughout the paper, we will use $v_A=B/\sqrt{4\pi\rho}$ for the Alfvén speed (which is subsequently used to calculate the Alfvén Mach number $\mathcal{M}_A$) versus $v_{Ai}$ for the ion Alfvén speed with $v_{Ai}=v_A/\sqrt{\chi}$.

Wave growth by the streaming instability competes with various damping mechanisms (ion-neutral damping, Landau damping, and turbulent damping, with the dominant mechanism depending on the CR energy and phase of the ISM \citep{Plotnikov_2021}). Coupling between the CRs and these self-excited waves work to reduce the bulk velocity of the CR streaming to within the ion Alfvén speed $v_{Ai}$ as equilibrium is reached between the streaming instability and wave damping. 

Specifically, Alfvén waves manifest as magnetic fluctuations that the CRs scatter off of. Therefore while in the fluid frame of reference the CRs are moving along magnetic field lines at the speed of light, they are streaming at approximately $v_{Ai}$ on average due to randomization from pitch angle scattering. These CRs are deemed ``self-confined". Alternatively, magnetic fluctuations can be created from other background turbulence, such as from turbulent energy cascades near the CR source sites.

S23 explores CR diffusion by combining simulations of background MHD turbulence with CR propagation experiments, in which streaming CRs also undergo diffusive acceleration driven extrinsically by turbulence in the background plasma advecting field lines on large (molecular cloud to galactic) scales. S23, like many others, assumes self-confinement, so the streaming speed of the CRs is fixed to the ion Alfvén speed. The S23 paper therefore aims to create a complete effective theory for CR transport that can be applied to large-scale simulations by incorporating components of advection, diffusion, and streaming at scales below the injection scale of turbulence.

S23 provides useful empirical fitting formulae which we utilize in this work to approximate various quantities of interest, as a function  of the Alfvén Mach number and the ionization fraction:
\begin{equation}
f(\mathcal{M}_{A}, \chi) = p_0\chi^{p1}+p_2\chi^{p_3}\Bigg{\{}\small \frac{\tanh[p_4\log_{10}(\mathcal{M}_A)-p_5] + 1}{2}\Bigg{\}}
\end{equation}
\noindent where the coefficients $p_i$ are given in Table 2 of S23 depending on the fit quantity chosen from {$\kappa_\parallel/\kappa_\perp$, $\kappa_\perp$, $u_\parallel/v_{\rm{str}}$}, where $\kappa_\parallel$ and $\kappa_\perp$ are the parallel and perpendicular diffusion coefficients, $u_\parallel$ is the mean streaming velocity in the direction along the local mean magnetic field, and $v_{\rm{str}}=B/\sqrt{4\pi\chi \rho}$ is the streaming speed in the form of the ion Alfvén speed. 

These fitting formulae yield values for the diffusion coefficient in dimensionless units, which are then rescaled with the factor $\ell_0^{\alpha-1} c_s \mathcal{M}$, where $\ell_0$ is the turbulence driving scale  $\ell_0=L_{\rm{inj}}=L_{\rm{min}}$, $\alpha$ is the superdiffusivity index ($\alpha=2$ corresponds to classical diffusion, while $\alpha<2$ corresponds to superdiffusive behavior), $c_s$ is the sound speed, and $\mathcal{M}$ is the sonic Mach number. To compare to our microscopic model we choose to use $\alpha=2$ in our main results but in Section \ref{sec:superdiff} compare the diffusion coefficients from classical versus superdiffusion using $\alpha=1.5$. 

Putting in $\mathcal{M}=\sigma_{\rm{1D}}/c_s$ and $\alpha=2$ yields units of $\ell_0\sigma_{\rm{1D}}$, where $\sigma_{\rm{1D}}$ is the 1-dimensional velocity dispersion of the simulation's unresolved (subgrid) turbulence. We therefore calculate $f(\mathcal{M}_A,\chi)$ for dimensionless $f = \kappa_\parallel/\kappa_\perp,~ \kappa_\perp$ and $\kappa_\parallel = (\kappa_\parallel/\kappa_\perp)\kappa_\perp$ for every grid cell and multiply them by $\ell_0\sigma_{\rm{1D}}$ to get our final model values in physical units. We invert the diffusion coefficient ratio to get $\kappa_\perp/\kappa_\parallel$ during our analysis as we felt it more physically intuitive to analyze a diffusion coefficient ratio between $0$ and $1$.

We calculate the ionization fraction $\chi$ for each simulation cell by linearly interpolating the cooling tables from \texttt{RAMSES} based on hydrogen number density $n_H$ and average ISM temperature $T$. We note that the cooling tables produced by \texttt{RAMSES} assumes a balance between heating and cooling with a fixed ultraviolent (UV) background that ignores any local interstellar radiation field. In reality, the UV radiation from nearby stars as well as heating from CR ionization should be included. These processes are absent from the MHD simulation we have post-processed for this work. The  formulation we utilize here is simpler, but we note that it may yield slight underestimates in $\chi$ and hence an underestimate of the diffusion coefficient in S23 which increases with $\chi$.

\subsection{ Critical turbulent diffusion coefficients}
\label{sec:critical}

For both our CR transport models (microscopic and macroscopic), there is a spatial scale below which CRs are confined by gas motions that dominate over CR diffusion. Diffusion that results only from turbulent mixing of the ISM with CRs frozen into the gas is associated with a turbulent diffusion coefficient. The general equation for this is $$\kappa_{\rm{turb}}=L\sigma/3.$$ 
for turbulence on a length scale $L$ with associated turbulent velocity dispersion $\sigma$. For any cell of our simulation we can therefore define $\kappa_{\rm{turb}}$ at the turbulence injection scale with $\kappa_{\rm{turb}}=L_{\rm{inj}}\sigma_{\rm{1D}}/3$, where again $L_{\rm{inj}}=L_{\rm{min}}$ and $\sigma_{\rm{1D}}$ is the velocity dispersion associated with subgrid (unresolved) turbulence. This is the turbulent diffusion coefficient associated with turbulence at subgrid scales, i.e. at wavenumbers higher in a turbulent cascade than we resolve. Alternatively, we may define a turbulent diffusion coefficient on resolved scales, with $ \kappa_{\rm{turb}}=L\sigma_i/3$ on some scale $L>L_{\rm{min}}$ and $\sigma_i$ is the resolved velocity dispersion in a given direction $\sigma_{x,y,z}$.

In both cases, $\kappa_{\rm{turb}}$ can be used as a metric for how the CR diffusion coefficient compares to the local turbulent flow: if $\kappa<\kappa_{\rm{turb}}$ (for $\kappa$ in either the perpendicular or parallel direction), advection of CRs by the gas will dominate over other transport processes on that scale. Later in Section~\ref{sec:kappaz} for example (see the grey curve in Fig.~\ref{fig:avgz}), we use a resolved $\kappa_z^{\rm{turb}}$ in the $z$-direction to estimate how confined CRs will be by the turbulence in the disk or in the corona.

In our MHD simulation, we obtain a minimum subgrid turbulent velocity dispersion around $\sigma_{\rm 1D} \simeq 10$~km/s for a minimum cell size $L_{\rm min} \simeq 10$~pc. This corresponds to an absolute floor value, called here the {\it cell} critical value,  $$\kappa_{\rm{turb, cell}} \simeq 10^{25} ~\rm{cm^2~s}^{-1},$$ below which CRs remain trapped within the corresponding computational cells. We furthermore define two critical scales for which the corresponding turbulent diffusion is likely to play an important role in confining CRs: the height of the galactic disk and the radius of the galactic corona. 
As shown below, the typical disk scale height is around $L\simeq 100~\rm{pc}$, with a disk velocity dispersion around $\sigma_{\rm{1D}}\simeq 10~\rm{km~s}^{-1}$, yielding a {\it disk} critical value $$\kappa_{\rm{turb, disk}}\simeq 10^{26}~\rm{cm^2~s}^{-1}.$$  
Turbulent motions in the galactic corona occurs on a significantly larger scale $L\simeq 1~\rm{kpc}$, with also a higher velocity dispersion $\sigma_{\rm{1D}} 
\simeq 100 ~{\rm km ~s}^{-1}$,  
yielding critical values in the corona of $10^{28}\rm{cm^2~s}^{-1}$. 
In our galaxy, for example, the aforementioned turbulent diffusion coefficient in the $z$ direction, $\kappa_z^{\rm{turb}}$, transitions from $10^{26}\rm{cm^2~s}^{-1}$ in the disk to between $3$ to $6\times 10^{28}\rm{cm^2~s}^{-1}$ between $z\sim 1$ to $10~\rm{kpc}$, as can be seen in Figure~\ref{fig:avgz}.  Therefore for our third critical value, the {\it corona} critical value, we choose a typical value of 
$$\kappa_{\rm{turb,corona}}\simeq 3\times 10^{28}~\rm{cm^2~s}^{-1}.$$
If the CR diffusion coefficient is below any of these 3 critical values, the CRs are likely to be trapped on each respective scale: in the cell, or in the disk, or in the corona. If they are trapped on any of these scales, CR energy density will accumulate and begin to exert pressure on their environment, therefore possibly having a dynamical effect on the galaxy’s evolution.

\subsection{Radial and vertical profiles: mass-weighting versus volume-weighting}
\label{sec:avg}

Throughout Section~\ref{sec:results} we will show vertical and radial profiles of quantities of interest. For this, we need to spatially average various properties of the galaxy (gas density, magnetic field, diffusion coefficients, etc.) in radial or vertical bins. For this, we rotate and translate each cell in our \texttt{RAMSES} snapshot using the python package \texttt{pynbody} to a reference frame centered on the halo center found using a shrinking-sphere method \citep{power2003}, with the gas disk's angular momentum vector pointed up in the positive $z$ direction. \texttt{pynbody} calculates the angular momentum vector using gas within a $10 ~\rm{kpc}$ sphere of center.

For reasons described below, we compute the average of a given quantity for all cells whose coordinates fall within each spatial bin (e.g. in a ring of radial width $\Delta R$ and height $\Delta z$) either i) weighted by cell volume, or ii) weighted by cell mass. We neglect the fraction of the volume of the cell that might fall outside of the bin.

When averaging a quantity within a given bin, it is important to satisfy fundamental conservation laws such as particle counts, gas mass, gas momentum or various energy densities.

For example, if one wants to plot the radial or vertical profile of the ionization fraction $\chi$, we need to perform a mass-weighted average within the bin defined by a discretization of the integral $\langle\chi\rangle_M=\int \chi \rho dV / \int \rho dV$. Indeed, dividing both the top and bottom by the proton mass, this integral form becomes simply $\langle\chi\rangle_M=N_e/N_H$, where $N_e$ is the total number of free electrons in the bin and $N_H$ is total number of atomic nuclei in the bin. This definition is connected to extensive physical quantities that satisfy by construction an integral form of a conservation law, here particle number conservation. We also have the nice property that the binned averages would approach the smoothed integral as the bin size becomes small. 

On the other hand, if one is interested in the magnetic field strength profile $B$, the natural choice is to perform a volume-weighted average of the magnetic energy density within the bin, defined as $\langle B^2 \rangle_V=\int B^2 dV / \int dV$ and then use $\langle B \rangle_V = \sqrt{\langle B^2 \rangle_V}$.  In this case, the extensive quantity is the magnetic energy and this defines the integral form we must use.

When either mass- or volume-weighting is the natural (or the physically motivated) choice for a given quantity, it is still useful to consider the other weighting scheme to highlight specific physical conditions in different phases of the ISM. A volume-weighted average will up-weight computational cells that have a larger volume, which, owing to the quasi-Lagrangian refinement strategy of \texttt{RAMSES}, corresponds to cells of lower relative gas density. Volume-weighting therefore essentially up-weights the warm (WNM) or hot (HIM) diffuse ISM. Weighting by mass conversely highlights regions of higher density like the cold neutral medium (CNM). Our different weighting methods can thus probe the physical conditions in different phases of the ISM (see Sect.~\ref{subsec:phase}).

Finally, we will show in the next sections that under most conditions, CRs can propagate freely out from molecular clouds and fill up the entire volume of the disk. In this case, it is more physically justified to compute the volume-average of the diffusion coefficient within the bin. If, on the other hand, CRs are trapped inside the dense ISM, a more meaningful approach is to use the mass-weighted average diffusion coefficient within each bin.

\section{Results}\label{sec:results}

\begin{figure*}
    \centering
    \includegraphics[width=1.75\columnwidth]{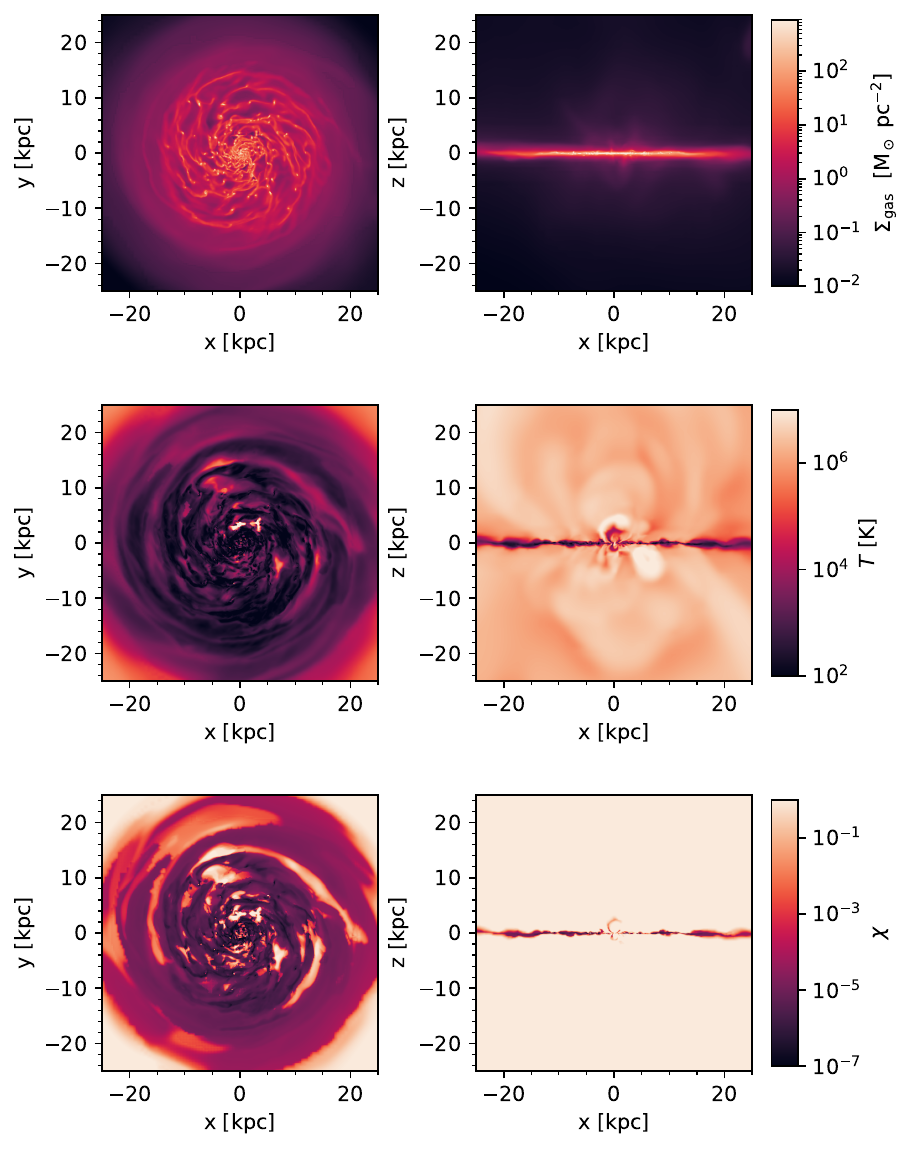}

    \caption{Spatial maps of physical properties in and around our galactic disk. The galaxy is an isolated MHD disk simulated using \texttt{RAMSES}. In each row the left panel shows a face-on view, the right an edge-on view, of the galaxy, coloured by a given quantity shown on each colorbar. From top to bottom we have gas surface density $\Sigma_{\rm{gas}}$, temperature $T$, ionization fraction $\chi$, one-dimensional turbulent velocity dispersion $\sigma_{\rm{1D}}$, magnetic field magnitude $B$, and Alfvén mach number $\mathcal{M}_{A}$. In general the disk is dominated in mass by cold and warm, neutral, dense gas compared to the warm and hot ionized gas that makes up the circumgalactic medium.}
    \label{fig:ingredientmaps}
\end{figure*}

\begin{figure*}
    \centering
    \includegraphics[width=1.75\columnwidth]{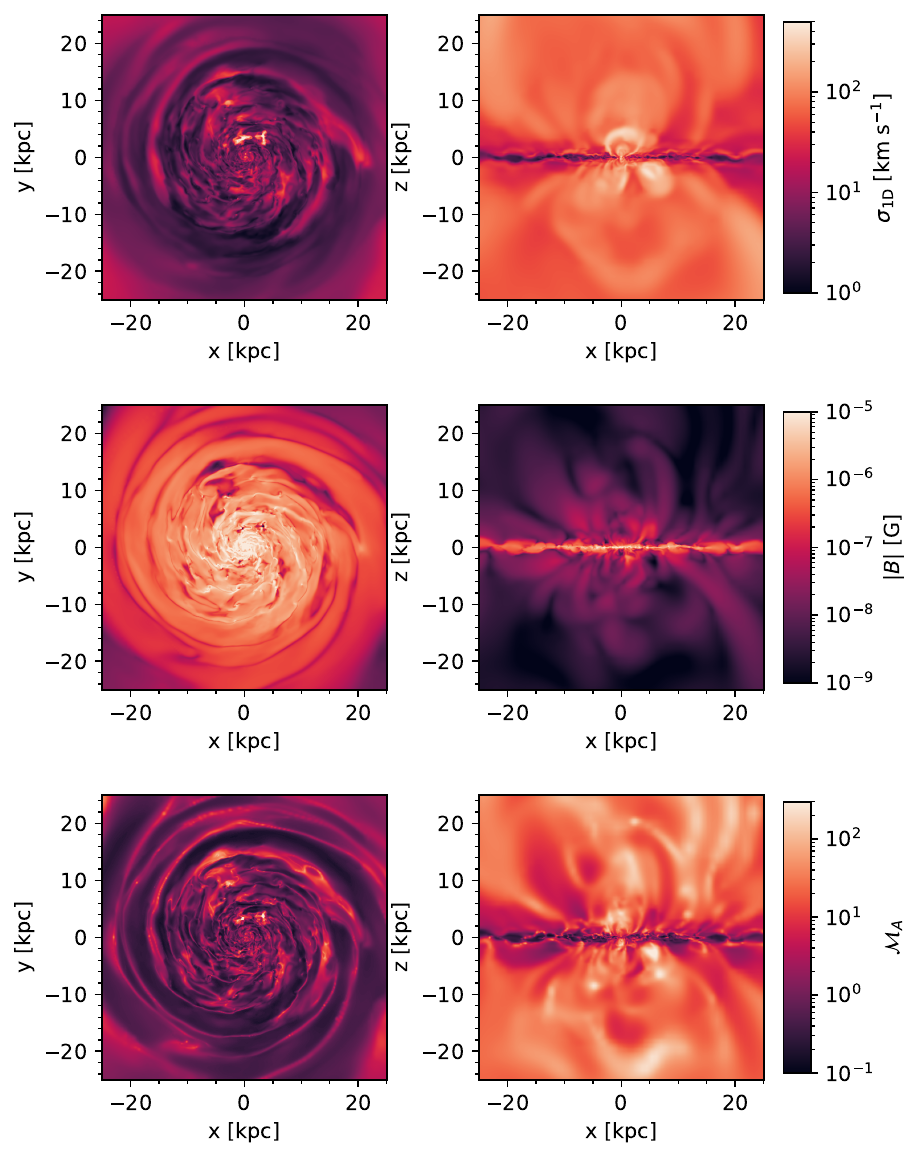}
    \caption{Continuation of Figure 1.}
    \label{fig:ingredientmaps2}
\end{figure*}

\begin{figure*}
    \centering
    \includegraphics[width=1.75\columnwidth]{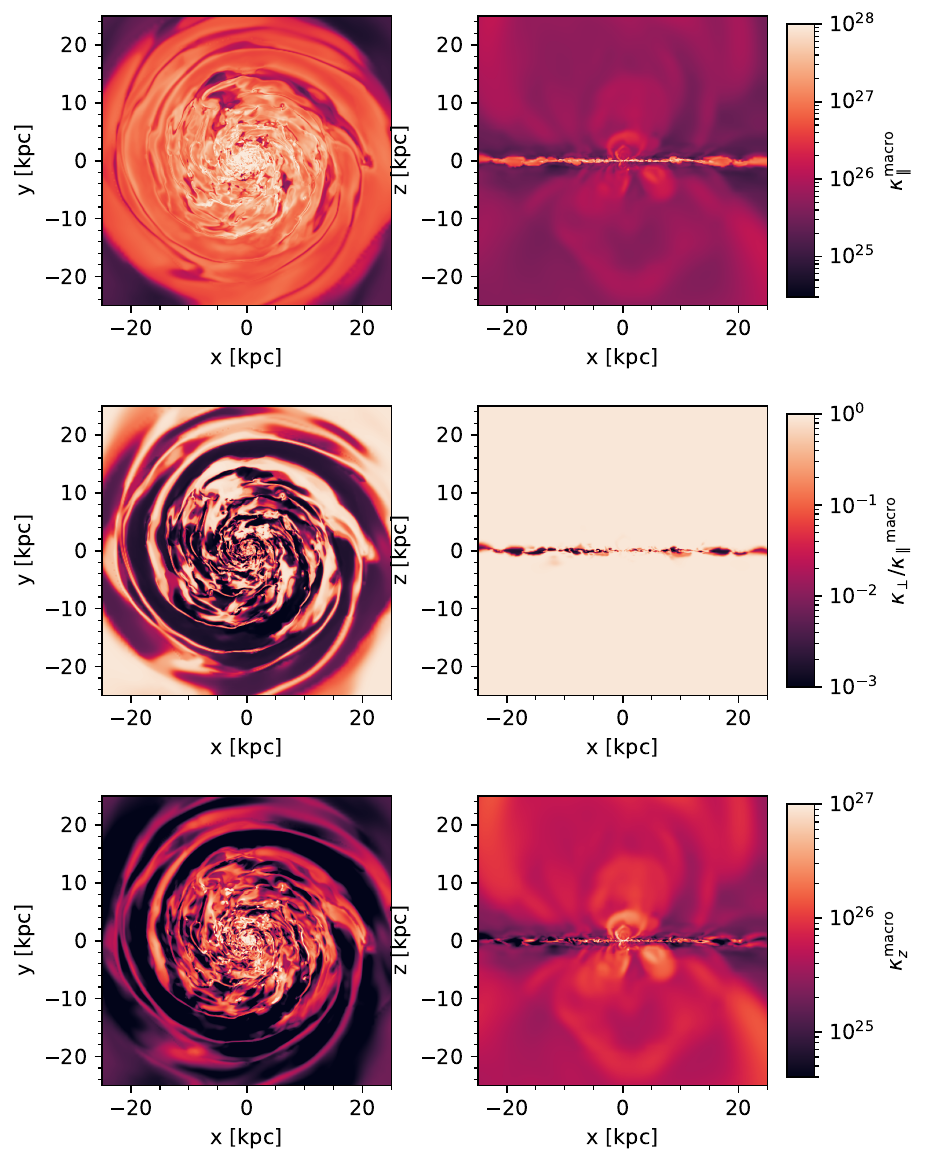}
    \caption{Spatial distributions of the diffusion coefficient for our macroscopic model. The three rows show, from top to bottom respectively, the diffusion coefficient parallel to the magnetic field, the ratio of the perpendicular to parallel diffusion coefficients, and the diffusion coefficient in the $z$-direction (out of the plane of the disk). The latter is derived in Sect.~\ref{sec:kappaz}. The highest values of $\kappa_\parallel$ and $\kappa_z$ can be found in the disk, especially near the galactic center, and decrease with radius and with height $z$. The CR diffusion is the most anisotropic ($\kappa_\perp/\kappa_\parallel \ll 1$) in this plane of the disk as well, and quickly becomes isotropic everywhere outside of the disk.}
    \label{fig:ingredientmaps3}
\end{figure*}

\begin{figure*}
    \centering
    \includegraphics[width=1.5\columnwidth]{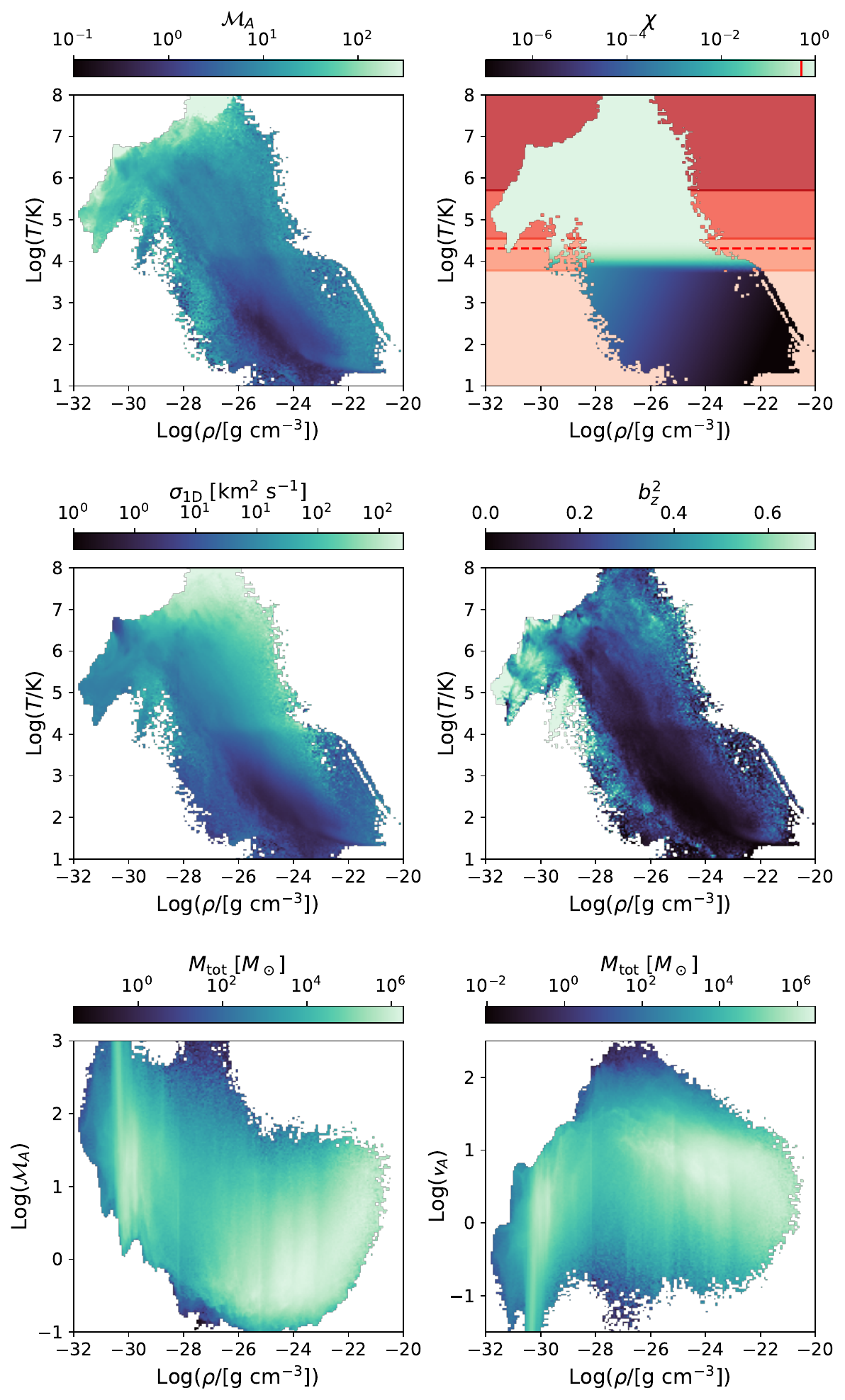}
    \caption{Density--temperature histograms binned in log-space. The histograms are coloured by the average value of a given parameter in that $\log\rho$--$\log T$ bin. The top two panels are $\mathcal{M}_{A}$ and $\chi$, and the middle left shows $\sigma_{\rm{1D}}$. All of the gas above $\log T\sim 20,000$ K is completely ionized, with a transition zone of partially ionized (the band of teal around $\log T\sim 10^4$), and then predominantly neutral at lower temperatures, with higher density being associated with lower neutral gas. Behind this histogram we have coloured the temperature regimes for the five phases we define in Table \ref{tab:phases}. All bins above the red dashed line has $\chi>0.5$; this boundary is also marked by a red line on the $\chi$ colorbar. The middle right panel shows a measure of the magnetic field orientation $b_z^2$. The bottom panels are slightly different, instead showing $\log\rho$ on the x-axis but $\log \mathcal{M}_{A}$ and $\log v_{A}$ on the y-axis instead. This rows's histograms are coloured by the total mass contained in the bin.}
    \label{fig:ingredienthists}
\end{figure*}

In this work we apply our model for the CR diffusion coefficient, in a post-processing scheme, to an isolated MHD disk simulated using the \texttt{RAMSES} code. The simulation has been presented in detail in \cite{Girma}, and we give a brief description in Section~\ref{sec:methods}. 
Both our microscopic and macroscopic models depend on the properties of the plasma in the disk (density, ionization fraction, temperature), the magnetic field (both in magnitude and orientation) and the strength of turbulence. In the following section we give the reader a sense for how these properties vary in our simulated galaxy. We then present our results in term of CR diffusion coefficients for both models.

\subsection{Galaxy properties}
\label{sec:ingredients}

The diffusion of CRs depends crucially on the phase of the ISM (or the CGM) in which they are evolving. Indeed, each gas phase shows very different properties such as temperature and ionization fraction. We categorize the ISM phases loosely based on Table 3 of \cite{kim2023}. This table includes nine gas phases defined by temperature and the relative abundance of molecular, atomic, and ionized hydrogen. We slightly simplify these definitions using only five phases based on the temperature $T$ and ionization fraction $\chi$, outlined in Table \ref{tab:phases}. For the cold ISM, we only define the cold neutral medium (CNM) (with $\chi<0.5$) since all the cold gas in our simulation is neutral. We also have the warm neutral medium (WNM), warm ionized medium (WIM), the warm-hot ionized medium (WHIM), and the hot ionized medium (HIM). The CNM, WNM, and WIM are in general confined to the disk, since outside of the disk (in the CGM), most of the gas is at high temperatures and ionized.

\subsubsection{Spatial distributions}

In Figures~\ref{fig:ingredientmaps} and \ref{fig:ingredientmaps2}, we show spatial maps of various galaxy properties. Each row is a face- and edge-on view of the galaxy, coloured by the density-weighed, line-of-sight average of the given property. From top to bottom we have the gas surface density $\Sigma_{\rm{gas}}$, temperature $T$, ionization fraction $\chi$, one-dimensional turbulent velocity dispersion $\sigma_{\rm{1D}}$, magnetic field strength $B$, and total Alfvén Mach number $\mathcal{M}_{A}$. 

As expected, the disk is dominated by the cold and warm neutral medium, with surface density ranging from 1 to 1000~$M_\odot~\rm{pc}^{-2}$ and $\chi \lesssim 0.1$. Dense gas clumps can be seen in the face-on surface density view, which are also cold and host active star formation.

The turbulent velocity dispersion ranges from 1 to 10~km/s in the disk. The magnetic field is highest in the disk, specifically in the inner $5$ kpc where $B$ reaches 10~$\mu G$ and above. This all contributes to $\mathcal{M}_{A}$ being of order unity throughout the disk. 

We also see that the CGM is dominated by hot ionized gas. In fact, gas is essentially completely ionized everywhere outside the disk as can be seen in the third row's edge-on Figure. The magnetic field far from the disk is very low (around a few $nG$ and below). We can see outflows of higher density gas coming out of the disk plane. These outflows have a velocity dispersion around 100 km/s and high temperatures above $10^6$~K. 

In the face-on temperature view in the second row, one can see two bright spots of high temperature gas just above the galaxy center. These are two supernovae expanding bubbles filled with hot, low-density, fast-moving and turbulent gas. 

In Figure~\ref{fig:ingredientmaps3}, we show spatial maps of a subset of the diffusion coefficient results for one of our models, the macroscopic model, as an illustrative example of how the diffusion coefficient varies spatially with other physical properties. We show the diffusion coefficient parallel to the magnetic field $\kappa_\parallel$, the ratio $\kappa_\perp/\kappa_\parallel$, and the diffusion coefficient out of the plane of the disk, $\kappa_z$, from top to bottom respectively. The derivation for and significance of $\kappa_z$ is discussed  in Sect.~\ref{sec:kappaz}. Both $\kappa_\parallel$ and $\kappa_z$ decrease with radius in the disk and with height away from the disk, and are relatively constant far from the disk, similar to the spatial trends in other physical properties like $|B|$ and $\sigma_{\rm 1D}$. The diffusion is isotropic ($\kappa_\perp/\kappa_\parallel\rightarrow 1$) everywhere outside the disk, similar to the spatial trends in ionization fraction $\chi$.

The same set of plots for our microscopic model would similarly show higher values in the plane of the disk than at greater $|z|$ values, but has both higher and more homogeneous $\kappa_\parallel$ and $\kappa_z$ values in the plane of the disk, and inverse radial trends (see Figure~\ref{fig:diffcoeff}).

\subsubsection{Phase space diagrams}

In Figure~\ref{fig:ingredienthists}, we show phase space diagrams of various key variables. In four of the six quadrants, we bin log of gas density $\rho$ on the x-axis and log of temperature $T$ on the y-axis, and create two dimensional histograms where each 2D bin is coloured by the average value of the depicted variable in that bin. 

We represent in these four quadrants the Alfvén Mach number $\mathcal{M}_{A}$, the ionization fraction $\chi$, the magnetic field orientation $b_z^2$ (defined below), and the 1D turbulent velocity dispersion $\sigma_{\rm{1D}}$. We expect in general that for low densities and high temperatures, the gas will be more turbulent, more ionized, and with a weaker and more tangled magnetic field, which is indeed what we see. 

As we can see in the top right panel, all of the gas with temperature above $20,000$ K is completely ionized, with a transition zone of partially ionized (the band of teal around $\log T\sim 4$), and then predominantly neutral at lower temperatures, with higher density being associated with more neutral (lower $\chi$) gas. 

In the background of this particular panel, we have coloured the regions between the temperature bounds of the phases of table~\ref{tab:phases} between cold, warm, warm-hot, and hot from light to dark red respectively. The delineation between the warm neutral and warm ionized medium is not discrete, but we have placed a dashed red line above which all cells have $\chi>0.5$. We have also placed a red line on the colorbar at $\chi=0.5$. The HIM (the darkest red) has densities below $\log_{10}(\rho)\sim -24$ for all temperatures; the CNM (the lightest red) has the highest densities, all above $\log_{10}(\rho)\sim -29$ and approaching $\log_{10}(\rho)\sim-20$. The WNM/WIM fall in between these two. These temperature regimes apply to all other panels as well. 

In the bottom two quadrants of Figure~\ref{fig:ingredienthists}, we have binned instead in gas density $\rho$ and Alfvén speed $v_{A}$ or Alfvén Mach number ${\mathcal M}_{A}$, coloured by the total mass contained in each bin. We see that a large portion of the mass is contained in the cold dense phase, with densities ranging from $\log_{10}(\rho)\sim -24$ to $-21$ ($1$ to $1000$~H/cc) and Alfvén Mach number ranging approximately from $0.5$ to $10$. This means that in the dense ISM, the magnetic field energy density is comparable or slightly weaker than the turbulent kinetic energy density.

\subsubsection{Radial profiles}\label{sec:radial}

\begin{figure*}
    \centering
\includegraphics[trim={0.7cm 0 0 0}, width=1.75\columnwidth]{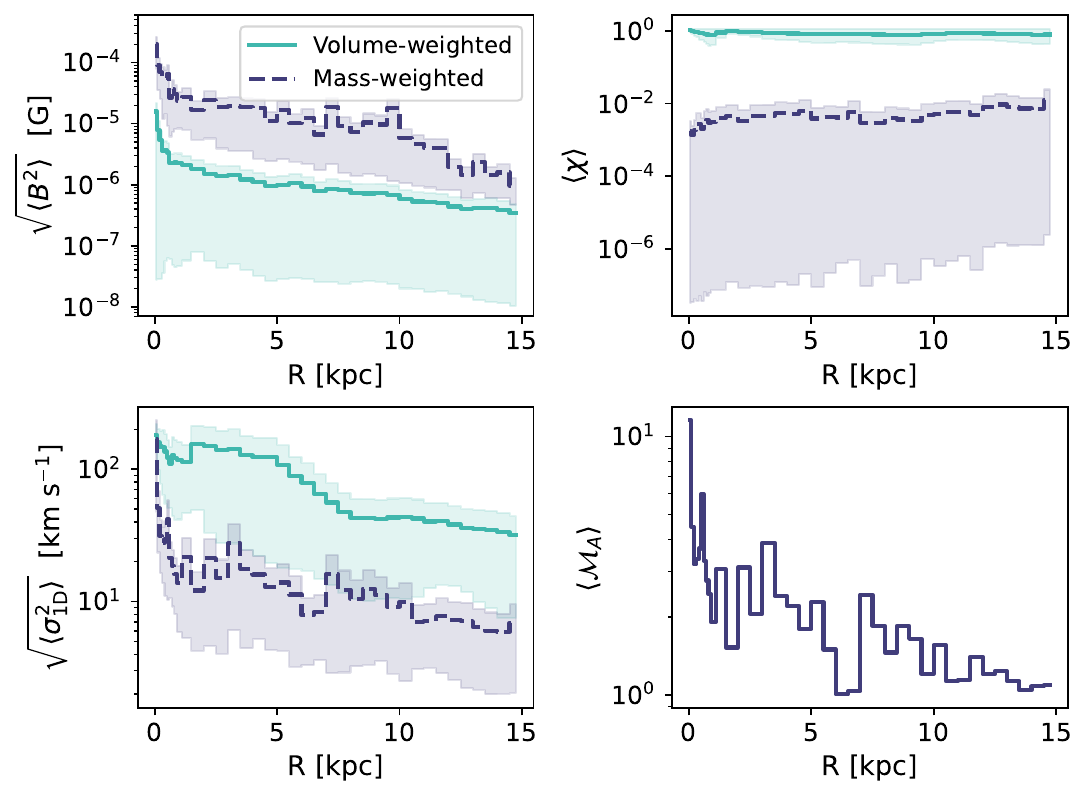}
    \caption{Average values of various parameters in the galactic disk as a function of radius. The top row shows the RMS average of magnetic field $B$ and ionization fraction $\chi$. The left lower panel shows the RMS average of the turbulent velocity dispersion $\sigma_{\rm{1D}}$. Each of the three panels shows a volume-weighted and mass-weighted average, with the former shown as a solid teal line and the latter as a dashed dark blue line. We also show a weighted confidence interval for the three panels for each weighting, calculated as discussed in Sect.~\ref{sec:radial}. Each weighting is associated with either a physically-motivated integral, or an average that highlights a particular phase of the ISM  (see Sections \ref{sec:avg} and \ref{sec:ingredients}). The bottom right panel combines the physically-motivated averages to calculate an average Alfvén mach number $\mathcal{M}_{A}$. $B$ and $\sigma_{\rm{1D}}$ both drop with radius, but the magnetic field has a higher mass-weighted average in contrast to the velocity dispersion having a higher volume-weighted average. The ionization fraction $\chi$ conversely increases with radius for the mass-weighted average and remains approximately constant and around $\chi\sim 1$ for the volume-weighted average. The Alfvén mach number also drops with radius from the center to the disk's edge by about an order of magnitude.}
    \label{fig:paramsradial}
\end{figure*}

In Figure~\ref{fig:paramsradial}, we have plotted the radial profiles of the magnetic field strength in the top left panel, the ionization fraction in the top right, the turbulent velocity dispersion in the bottom left and the Alfvénic Mach number in the bottom right. In each panel, we show both the mass-weighted and the volume-weighted average in each radial bin. The bin averages are computed within $z = \pm 2$ kpc, with a constant radial bin size $\Delta R=0.1$ kpc, out to $R=1$ kpc and then $\Delta R=0.5$ kpc out to $R=15$ kpc. The ring volume is computed using $\Delta V = \Delta z \pi (R_2^2-R_1^2)$. The bin radial coordinate is computed using the midpoint $R=(R_1+R_2)/2$).

As discussed in Sect.~\ref{sec:avg}, the two different averaging schemes give us access to different physical process and different gas phases in the galaxy. In contrast to the volume-weighted magnetic field strength, the mass-weighted magnetic field strength does not have a proper physical interpretation (i.e. integrating $B^2\rho dV$ does not correspond to any known physical property). It instead serves to highlight the typical magnetic field conditions in the cold dense ISM (see Sect.~\ref{sec:avg}). We see that the mass-weighted magnetic field strength is higher than the volume-weighted value. This is due to the higher magnetic field strength in the dense clumps owing to its adiabatic compression, as we saw in the spatial plots of Figure~\ref{fig:ingredientmaps}. 

In the case of the ionization fraction $\chi$, the mass-weighted average is associated to a proper conservative integral form. Indeed, the mass-weighted average is akin to summing the total number of free electrons $N_e$ in the ring and then dividing by the total particle number in the bin $N_{\rm H}$ to get an average ionization fraction. The mass-weighted average remains below $\sim 10^{-2}$ across the disk. The volume-weighted average, on the other hand, highlights the conditions in the warm/hot ionized mediums, with the average ionization fraction sitting around $\chi\sim 1.0$. 

The mass-weighted turbulent velocity dispersion is equivalent to integrating the total turbulent kinetic energy in a bin and then finding the average specific kinetic energy. We see that $\sigma_{\rm{1D}}$ drops from $\simeq 100$~km/s in the very center of the galaxy to $\sigma_{\rm{1D}}\sim 10$~km/s across most of the disk. The volume-weighted velocity dispersion is significantly higher than the mass-weighted average. This highlights the conditions in the warm/hot ISM with stronger but subsonic turbulence. 

Finally, the fourth quadrant combines the other variables to compute the bin-averaged Alfvén Mach number $\mathcal{M}_{A}$. For this, we use a combination of the physically-motivated averages: We use the mass-weighted velocity dispersion $\sigma_{\rm{1D}}$, the volume-weighted field strength $B$, and the volume-weighted average gas density $\rho$, using the following equation:
\begin{equation}
\langle \mathcal{M}_{A} \rangle =  \sqrt{\frac{\langle\sigma^2_{\rm{1D}}\rangle_M 4\pi \langle\rho\rangle_V}{\langle B^2\rangle_V}}
\end{equation}
The radicand is equivalent to dividing the average turbulent kinetic energy by the average magnetic energy over the whole bin which, with the prefactor $4\pi$, is equivalent to $\mathcal{M}_{A}^2$. Using this definition, we see that $\mathcal{M}_{A}$ reaches $10$ at the center of the disk, but remains close to $1$ throughout most of the disk.

To give a sense of the underlying parameter distribution, we also calculate a weighted 68\% confidence interval with respect to the mean. We sort the data in a given bin by its absolute distance from the mean, and then find the minimum distance from the mean such that the sum of weights within that distance is 68\% of the total weight. We plot this interval as a shaded band around each profile. We do not calculate such a value for the Mach number since it is a quantity based on average conditions in a region and so a variance is not meaningful.

\begin{figure}
    \centering
\includegraphics[width=\columnwidth]{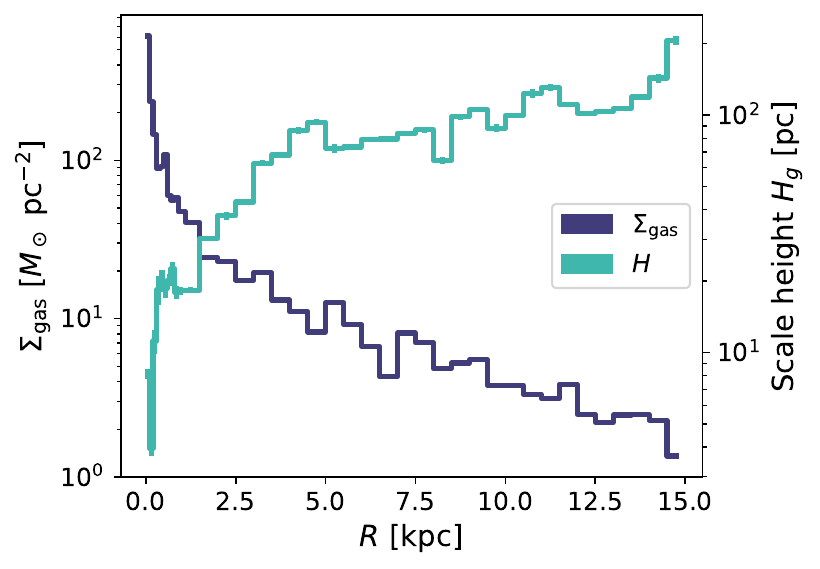}
    \caption{Gas surface density $\Sigma_{\rm{gas}}$ and scale height $H_g$ as a function of radius. The bin size increases from $\Delta R=0.01$ kpc to $\Delta R = 0.05$ within and outside of $R=1$ kpc respectively. $\Sigma_{\rm{gas}}$ drops with radius which consequently causes the scale height to increase. The scale height is found by fitting the cumulative distribution function of mass with height as described in Section \ref{sec:diffcoeff}.}
    \label{fig:sigma_H}
\end{figure}

In Figure~\ref{fig:sigma_H}, we show the radial profiles of the gas surface density $\Sigma_{\rm gas}$ and the vertical scale height $H_g$.  Both averages are calculated with the same bin sizes as the previous figure. The gas surface density falls exponentially from a value of $\sim 600~M_\odot~\rm{pc}^{-2}$ in the center, to a value between 1 and $10~M_\odot~\rm{pc}^{-2}$ in the rest of the disk. 

To compute the disk scale height $H_g$, for each radial bin we create a logarithmic binning in $|z|$ out to $|z|=2~\rm{kpc}$ and we sum up the gas mass cumulatively. We then normalize by total mass to get the vertical Cumulative Distribution Function (CDF). Assuming a mass density profile that decreases exponentially with height, we can fit the CDF using the function $f(z) = 1 - e^{-|z|/H_g}$. We do this using python's \texttt{scipy} curve fitting function to get values for $H_g$ and an error $\sigma(H_g)$ for each $R$ bin. 

The scale height is quite small, with $H_g \le 10$~pc in the nuclear region of the galaxy, but remains close to $100$~pc throughout most of the galaxy. At $R > 15$~kpc, $H_g$ begins to increase again as we enter the CGM. We use a value of $H_g\simeq 100$~pc as our representative disk scale height for the rest of our analysis.

\subsection{CR Diffusion Coefficient}

We now compute the corresponding diffusion coefficients for three models: our microscopic model (YL08), macroscopic model (S23) and an additional model with super-diffusion (S23). We will see that these three models result in very different (by orders of magnitude) values of their corresponding effective diffusion coefficients. Once we have computed the local diffusion coefficients, we then compute their average vertical component in each model. This will allow us to model the vertical transport of CRs flowing out of the galaxy. 

As a note of caution, we would like to stress that we are not modelling here any backreaction of CRs onto their host galaxy. First, a proper treatment of CR transport should include a proper treatment of advection in case the diffusive wave speed is locally lower than the gas advection speed. Second, if CRs remain trapped in the disk, their corresponding energy density should increase, possibly up to a point where the CR pressure would impact the vertical pressure support and the disk would expand. This could also raise the average ionization fraction, in turn raising the diffusion coefficient. 

These processes are missing from our current post-processing approach. Assembling a complete picture for CR transport that incorporates all factors and reproduces observed values is a challenge faced by all studies of CRs due to the vast range in length scales one would need to resolve (see a discussion of this in, e.g., \citealt{Kempski22}). We will explore these details further in a companion paper of this study.

\subsubsection{Diffusion coefficient parallel and perpendicular to the magnetic field}
\label{sec:diffcoeff}

\begin{figure*}
    \centering
    \includegraphics[trim={1.5cm, 0cm, 2.75cm, 0}, width=2\columnwidth]{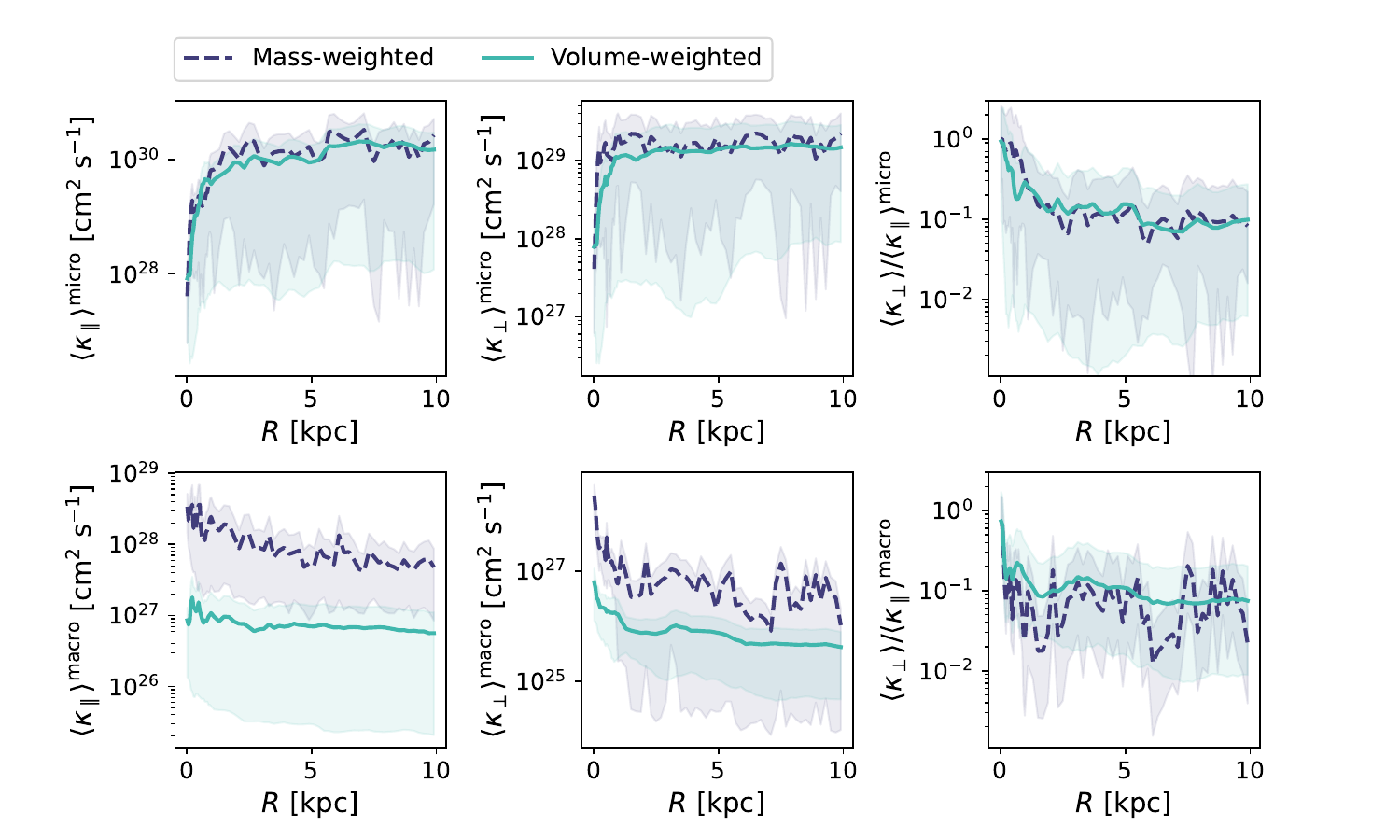}
    \caption{CR diffusion coefficient for our two models. The diffusion coefficient parallel and perpendicular to the magnetic field, and the ratio of the two, is shown. The top row shows the macroscopic model, the bottom row the microscopic model. Two different averaging methods are shown: the mass-weighted average is shown as a dashed dark blue line, and the average weighted by cell volume is in solid teal. Our weighted 68\% confidence interval measure is shown as colored bands around each profile. The two models have $\kappa$'s that differ by about three orders of magnitude, between $10^{26}$ and $10^{30}$ $\rm cm^2~s^{-1}$, and have a ratio $\kappa_\perp/\kappa_\parallel < 1$ over the majority of the galactic disk.}
    \label{fig:diffcoeff}
\end{figure*}

In Figure~\ref{fig:diffcoeff}, we show the average diffusion coefficient parallel ($\kappa_\parallel$) and perpendicular ($\kappa_\perp$) to the magnetic field, calculated as described in Sections \ref{sec:modelmicro} for the microscopic model and \ref{sec:modelmacro} for the macroscopic model, as a function of radius from the center of the disk. We use the superscripts ``\textit{micro}" and ``\textit{macro}" to denote each model.  The top row shows the macroscopic model, the bottom row the microscopic model. The volume-weighted average is shown with the solid teal line, and mass-weighted with a dashed dark blue line (see Sect.~\ref{sec:avg} for discussion on different weightings). We bin in rings with $\Delta R=0.05~\rm{kpc}$ out to $R=1.0~\rm{kpc}$ and then $\Delta R=0.2~\rm{kpc}$ out to $R=10~\rm{kpc}$, with a ring thickness of $\Delta z = 2~\rm{kpc}$ (that is, including cells with coordinates $z \leq \pm 1~\rm{kpc}$). We also calculate a 68\% confidence interval in each bin using the same process as for Figure~\ref{fig:paramsradial} discussed in Sect.~\ref{sec:radial} and plot this as colored bands.

The two models yield $\kappa$ values that differ by almost three orders of magnitude. The macroscopic model's diffusion coefficient decreases from the center of the disk outwards by an order of magnitude in the perpendicular direction, and by only a factor of 2 for the parallel direction. The microscopic model conversely \textit{increases} from the disk outwards, by an order of magnitude in the perpendicular direction and by two orders of magnitude in the parallel direction. In both models, $\kappa_\parallel$ asymptotes to a relatively constant value after  $R = 1~\rm{kpc}$.  The microscopic model reaches  $\langle\kappa_\parallel\rangle^{\rm{micro}} \simeq 10^{30} ~\rm{cm^2~s^{-1}}$ and $\langle\kappa_\perp\rangle^{\rm{micro}} \simeq 10^{29} ~\rm{cm^2~s^{-1}}$.  The macroscopic model, on the other hand, reaches $\langle\kappa_\parallel\rangle^{\rm{macro}}\simeq 10^{27}~\rm{cm^2~s^{-1}}$ and $\langle\kappa_\perp\rangle^{\rm{macro}}\simeq 10^{26} ~\rm{cm^2~s^{-1}}$, with exact values depending on the weighting scheme adopted.

We also show in the righthand column of Figure~\ref{fig:diffcoeff} the ratio of the average diffusion coefficients, as a proxy for how trapped in the plane of the disk the CRs will be; the magnetic field will, in general, lie in the plane of the disk, so if $\langle\kappa_\perp\rangle/\langle\kappa_\parallel \rangle \ll 1$, the CRs are likely to be trapped in the disk. Both models have a diffusion coefficient ratio $\langle\kappa_\perp\rangle/\langle\kappa_\parallel \rangle$ that decreases with radius. For both models (but more pronounced for the macrocopic model), the mass-weighted average is often lower than the volume-weighted average. This is because higher-volume (low-density) cells generally have higher Mach numbers and therefore more isotropic CR diffusion. These regions of free-streaming CRs contribute to the efficient CR mixing that we expect in the disk of the galaxy. For both models we find that $\langle\kappa_\perp\rangle/\langle\kappa_\parallel \rangle \simeq 0.1$ throughout the disk, implying that the CRs are mostly trapped in the galaxy. This also justifies our basic assumption in what follows that the CR energy density is mostly uniform within the disk - since CRs are able to escape their injection sites and move fast enough in the disk to homogenize their distribution - and that CR transport mostly occurs in the vertical direction.

The radial trends in $\kappa_\parallel$ and $\kappa_\perp$ can be explained using our discussion in Section~\ref{sec:ingredients}. We observe i) a radial decrease in gas surface density, ii) a radial decrease in Alfvén Mach number, but always above $1$, iii) a roughly constant average ionization fraction. The majority of the disk gas is trans-Alfvénic, resulting, for the microscopic model, in a roughly isotropic diffusion coefficient scaling as $\mathcal{M}_A^{-2}$, therefore increasing with radius when the average Mach number decreases with radius. For $\mathcal{M}_A>1$, the macroscopic model's diffusion coefficient is quite sensitive to the ionization fraction. The average ionization fraction remains roughly constant throughout the disk, explaining why the radial decrease of the diffusion coefficient is relatively modest. 

Typical values of diffusion coefficients inferred from observations, both for the disk \citep{3} and for the CGM \citep{1a,1b}, are typically in the range $10^{28}$ to $10^{29}~\rm{cm^2~s^{-1}}$ at $1$ GeV energies. These values may be altered depending on factors like magnetic rigidity (see e.g. \citealt{rigidity}), an effect which we do not explicitly explore in this work, or CR acceleration efficiency (see e.g. \citealt{Pandey2024}). The values that we find in our model are comparable to these previous studies, although are higher (lower) for the microscopic (macroscopic) model.

\subsubsection{Superdiffusion coefficient}
\label{sec:superdiff}

The accepted theory for transport of CRs is a combination of advection by the background gas, streaming along magnetic field lines, diffusing in the direction parallel to the magnetic field, and diffusing perpendicular to the magnetic field via field line wandering in the form of a random walk. However, numerical studies of anomalous diffusion have found that CR transport may be superdiffusive instead. This is a consequence of the divergence of magnetic field lines. In the ISM, there are turbulent eddies on a range of scales due to the turbulent cascade in the plasma, meaning as a CR travels over a given distance it will encounter field lines whose separation is accelerating (often noted to be analogous to Richardson's diffusion - see \citealt{Richardson1926}). Therefore on scales at or below the CR gyroradius where field line wandering is occurring, superdiffusive transport may arise \citep[see Sections 2.3 and 3 of][and references therein]{LazReview}.

These studies typically identify superdiffusion in the direction perpendicular to the magnetic field, and on small scales of order the CR gyroradius. Conversely in \cite{Sam22}, they find that their resulting CR transport can best be described as superdiffusive in \textit{both} the parallel and perpendicular direction, on the large scales that they consider. Recall that their study focusses on streaming CRs undergoing effective diffusion through advective motion. Further recall that the scaling for their resulting diffusion coefficient is $\ell_0^{\alpha-1} c_s \mathcal{M}=\ell_0^{\alpha-1}\sigma_{\rm 1D}$ for superdiffusivity index $\alpha$ (see Sect.~\ref{sec:modelmacro}). For our macroscopic model, we used a superdiffusivity index of $\alpha=2$ such that both $\kappa^{\rm{micro}}$ and $\kappa^{\rm{macro}}$ had units corresponding to classical diffusion. However in S23, all best fit values for $\alpha$ lay between $1.4 < \alpha < 2$, indicating superdiffusivity across their explored parameter space ($\alpha<2$ for superdiffusion), with most trials with $\mathcal{M}_{A}>1$ having $\alpha\simeq 1.5$. They do note one exception that below $\mathcal{M}_A<1$ and $\chi<0.01$, $\alpha\rightarrow 2$ in the direction parallel to the magnetic field. 

It is possible that a superdiffusive description of CR transport for our macroscopic model might yield diffusion coefficients more comparable to the microscopic model or empirical values. We therefore use the fitting formulae from the macroscopic model with $\alpha=1.5$ to calculate superdiffusive coefficients $\kappa_\parallel^{\rm sup}$ and $\kappa_\perp^{\rm sup}$. However, as we can see from the scaling $\ell_0^{\alpha-1}\sigma_{\rm 1D}$ of S23, $\alpha=1.5$ corresponds to diffusion coefficients with units of $\rm cm^{3/2} ~s^{-1}$. Our other two models follow classical diffusion with units of $\rm cm^2~s^{-1}$. To compare across models, we create an effective diffusion coefficient by multiplying $\kappa^{\rm{sup}}$ by the square root of a chosen length scale, $\kappa \equiv \sqrt{L}\kappa^{\rm{sup}}$.

In the next section we calculate a diffusion coefficient $\kappa_z$ in the $z$-direction. For this direction the obvious choice of length scale is the $z$-coordinate of a given cell, i.e.e we multiply $\kappa_z^{\rm sup}$ by $\sqrt{|z|}$. This means that the effective diffusion rate increases with distance from the midplane where the CRs are injected (assuming most CRs are injected into the galaxy by supernovae that in general lie in the disk, which is also very thin). This implies that there is a distance above which the transport speed for superdiffusion overtakes that of classical diffusion, and therefore it will take a CR less time to move a given distance than it would for classical diffusion (for an illustration of this see Figure 13 in S23). 

\begin{figure*}
    \centering
    \includegraphics[trim={1.5cm, 0cm, 2.75cm, 0}, width=2\columnwidth]{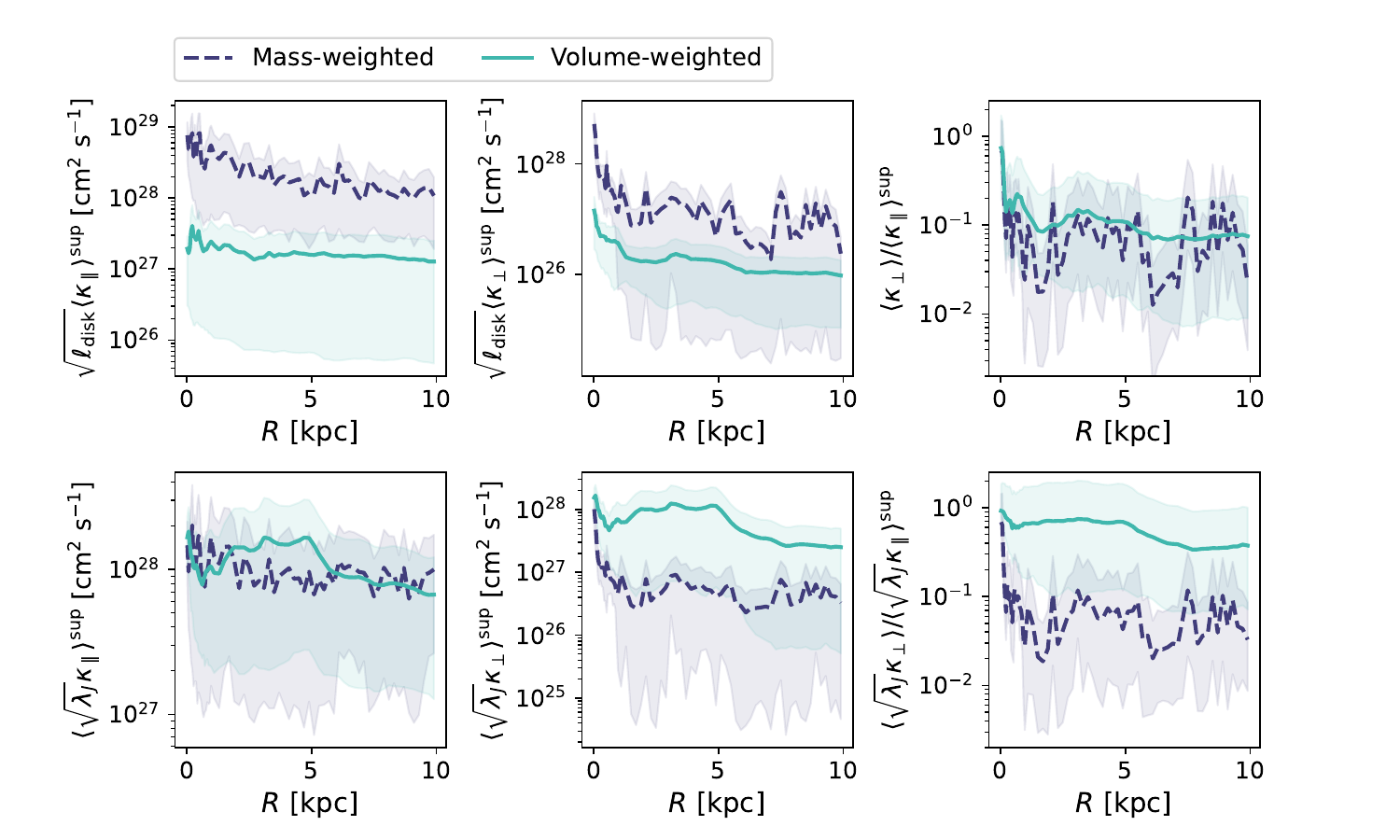}
    \caption{Effective diffusion coefficients calculated from the superdiffusive coefficient of Section \ref{sec:superdiff} using $\kappa=\sqrt{L}\kappa^{\rm{sup}}$ for length scale $L$. From left to right, the columns show $\kappa_\parallel$, $\kappa_\perp$, and the ratio of the two averages respectively. Mass-weighted is shown in dashed dark blue and volume-weighted in solid teal. The top row uses $L=\ell_{\rm disk}=100$ pc, and the bottom row uses the thermal jeans length $L=\lambda_J$ of each cell before averaging. Using $\lambda_J$ produces values falling in a narrower $\kappa$ range than in the upper panel. Our weighted 68\% confidence interval measure is shown as colored bands around each profile.}
    \label{fig:kparperpsup}
\end{figure*}

For $\kappa_\parallel$ and $\kappa_\perp$ the choice of length scale is not as obvious. We first try using a constant scale for all cells to see how fast CRs are diffusing once they have traveled distance $L$. For this we choose $L=\ell_{\rm disk}$,  where $\ell_{\rm disk}=100~\rm{pc}$ is the scale height of the disk, mimicking how fast the CRs would be diffusing when they escape a star forming region in the galaxy. Alternatively, we can use the scale $L=\lambda_J$ where $\lambda_J=(15k_BT/4\pi G\mu \rho)^{1/2}$ is the thermal Jeans length. This length scale can be thought of as a local gas scale height depending on each given phase, larger for warm/hot diffuse gas than for cold dense gas. 

We show examples of averaging for each of these cases in Figure~\ref{fig:kparperpsup}. The effective diffusion coefficients parallel and perpendicular to the magnetic field are shown in the first and second columns respectively, and the ratio of the two averages in the rightmost column. We use the same radial binning and line styles as the previous figure, and again show weighted confidence interval bands. We use the length scale $\ell_{\rm disk}$ in the top row, and $\lambda_J$ in the bottom row. Note that for the $\lambda_J$ case, we multiply every cell by its local $\lambda_J$ value \textit{before} averaging all cells for a given bin in $R$, that is $\langle\kappa_{\parallel,\perp}\rangle=\langle\sqrt{\lambda_J}\kappa_{\parallel,\perp}^{\rm sup}\rangle$. There are two SNe in the galaxy at approximately $R=3$ and $R=5$ kpc that create two very subtle bumps in the volume-weighted lines because of the low-density, high-temperature gas surrounding them. We find that recalculating these radial averages with the SNe data cut out does not change our results significantly.

In the upper panel that uses $\ell_{\rm{disk}}$, the mass- versus volume-weighted averages differ by about an order of magnitude for a given direction (parallel or perpendicular). In the lower panel that uses $\lambda_J$, the two weightings for $\kappa_\perp$ differ by more than an order of magnitude, but interestingly the volume- and mass-weighted curves largely overlap for $\kappa_\parallel$. These two curves, as well as the volume-weighted $\kappa_\perp$, remain around $10^{28}~\rm cm^2 s^{-1}$ across much of the disk; using $\lambda_J$ appears to somewhat unify the $\kappa$ averages by taking gas phase into account. 

Superdiffusion implies that the farther the CR travels, the faster it is diffusing. For some distance traveled $d$, by our units the ratio $\kappa/\kappa^{\rm macro}=\sqrt{d}\kappa^{\rm sup}/\kappa^{\rm macro}=\sqrt{d/\ell_0}$. Therefore for large enough $d$ such that $\sqrt{d/\ell_0}>1$, $\kappa^{\rm sup}$ is a ``boosted" version of $\kappa^{\rm macro}$. This is interesting because i) this can raise $\kappa$ to values closer to empirical estimates, and ii) at large enough distances, the CRs may then be diffusing fast enough to escape regions that they would otherwise be trapped in for $\kappa^{\rm macro}$, eventually free streaming away to infinity above some height. Interestingly, it appears that the effective diffusion coefficients we show in Figure~\ref{fig:kparperpsup} are indeed closer to empirical estimates (which are usually between $\sim 10^{28}~\rm cm^2 s^{-1}$ and $\sim 10^{29}~\rm cm^2 s^{-1}$) in the Milky Way than for $\kappa^{\rm macro}$. 

The fact that the superdiffusive model produces more similar values than the macroscopic model to observations presents a computational challenge, since how to implement a superdiffusive diffusion coefficient into the energy conservation equations in \texttt{RAMSES} that defines the self-consistent simulation is not obvious - from the outset the units are not compatible to make a physically meaningful implementation of $\kappa^{\rm sup}$. For now, we remain confined to the classical diffusion cases when implementing this subgrid model into simulations. However, how well each of our models reproduces CR observations will be further revealed after a full implementation of the subgrid model, and we cannot make conclusions about whether or not the superdiffusive model is actually most accurate. Galaxies will respond to the CR transport energetically and possibly dynamically as the the CR population and the rest of the galaxy components (gas, dust, stars, magnetic fields, etc.) co-evolve. The gas properties may change in response to induced CR pressure, which is turn will modify the local diffusion coefficient. Furthermore, gas advection will serve as an additional mode of CR transport since magnetic fields will be advected along with the gas. 

\subsubsection{Diffusion coefficient in the z-direction}
\label{sec:kappaz}

\begin{figure*}
    \centering
    \includegraphics[width=2\columnwidth]{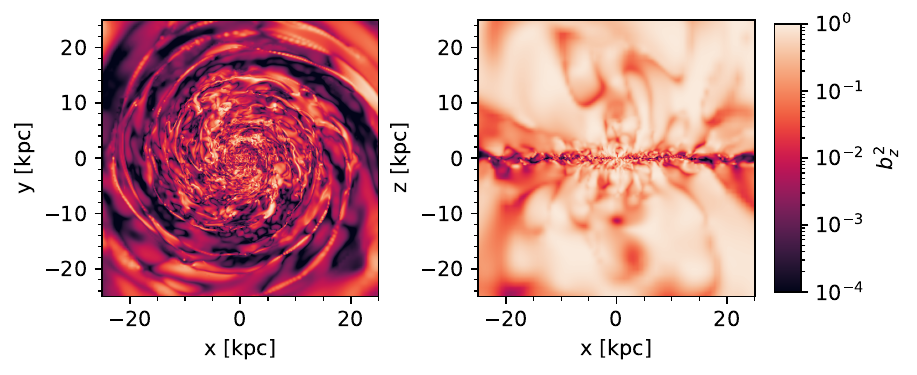}
    \caption{Spatial distribution of $b_z^2$ where $b_z$ is the $z$-component of the magnetic field. The left panel is the galaxy face-on, the right panel edge-on. In the disk, in general $b_z^2\ll 1$, whereas in the galactic corona there is a significant magnetic field orientation in the $z$-direction.}
    \label{fig:bzavgspatial}
\end{figure*}

As discussed in the introduction, when CR protons interact with particles in the ISM they produce neutral pions that quickly decay into gamma ray pairs. We can use the observed gamma ray luminosity from star forming galaxies in the GeV-TeV range to estimate the CR energy density $U$. Calculating $U$ from gamma ray emission or vice versa requires quantifying the calorimetric fraction $f_{\rm{cal}}$ of the disk, i.e. the fraction of CRs emitted via supernova shocks whose fate is to collide with particles in the ISM and thus feed back into the galaxy environment. We can also write this as $f_{\rm{cal}}=1-f_{\rm{esc}}$, where $f_{\rm{esc}}$ is the escape fraction of the galaxy, i.e. the fraction of CRs produced that leak out of the galaxy. 

To investigate the CR calorimetric fraction and gamma ray luminosity of the disk, we must first quantify the flow of CRs out of the plane of the disk. We have so far defined diffusion coefficients parallel and perpendicular to the local magnetic field. While the majority of the magnetic field will indeed be oriented in the plane of the galaxy (the $x$-$y$ plane), we cannot assume that the $z$-direction diffusion coefficient $\kappa_z$ is equal to $\kappa_\perp$, as there may be contributions by $\kappa_\parallel$ due to the $z$-component of the magnetic field vector. 

We thus must calculate $\kappa_z$ in a more formal manner. Anisotropic diffusion can be modelled using an anisotropic flux function written as
\begin{equation}
    {\bf F}_{\rm CR} = - {\mathbb D} \nabla U
\end{equation}
where the diffusion matrix writes
\begin{equation}
    {\mathbb D} =  \kappa_{\rm iso} {\mathbb I} + \kappa_{\rm ali} {\bf b} \otimes {\bf b}
\end{equation}
where the isotropic component $\kappa_{\rm{iso}}=\kappa_\perp$ and the magnetic field-aligned component is $\kappa_{\rm ali}=\kappa_\parallel-\kappa_\perp$ \citep[see e.g.][and references therein]{DC2016}. The unit vector ${\bf b}=(b_x,b_y,b_z)$ is pointing in the direction of the local magnetic field. In component form, the flux in the z-direction just writes
\begin{equation}
F_{{\rm CR},z} = - \kappa_{\rm ali} \left( b_xb_z \partial_x U + b_y b_z \partial_y U + b_z^2 \partial_z U \right) - \kappa_{\rm iso} \partial_z U
\end{equation}
Based on our previous discussion, we now assume that the CR energy density in the galaxy is approximately invariant along R, so $\partial_x U=\partial_y U \simeq 0$. This approximation is valid as long as CRs are moving at high enough speed to homogenize their distribution in the disk. It is also strengthened by the fact that the gaseous disk is really thin in our case with $R_d \simeq 34 H_{\rm g}$. This property is also supported by observations in the Milky Way \citep{Stepanov2014}. 
This leads to the final simple result:
\begin{equation}
F_{\rm{CR},z}=-((\kappa_\parallel-\kappa_\perp)b_z^2+\kappa_\perp)\partial_zU
\label{eq:diffz}
\end{equation} 
Comparing this to a flux equation with a single diffusion coefficient, $F_{{\rm CR},z}=-\kappa_z\partial_z U$, gives us an effective diffusion coefficient in the $z$-direction: 
\begin{equation}\label{eq:kappaz}
\kappa_z = (\kappa_\parallel-\kappa_\perp)b_z^2+\kappa_\perp = \kappa_\parallel \sin^2 \theta + \kappa_\perp \cos^2 \theta
\end{equation}
where $\theta$ is traditionally called the ``tilt angle'' of the magnetic field with respect to the galactic plane. We have thus $b_z = \sin \theta$. The first term in Equation~\ref{eq:kappaz} is an ``extra" diffusion coefficient contribution in the $z$-direction that comes from the magnetic field orientation not lying perfectly in the plane of the galaxy and so of course depends on $b_z$ or $\theta$. From this equation we conclude that the orientation of the galaxy's magnetic field is key in determining the strength of CR transport out of the plane of the disk. 

In Figure~\ref{fig:bzavgspatial} we show the spatial distribution of $b_z^2$ in our isolated, star-forming galaxy. The left panel shows the galaxy face on, the right panel edge-on. $b_z^2$ is lower in the galactic disk than in the corona, as expected for a magnetic field oriented mainly in the galactic plane. 

\begin{figure}
    \centering
    \includegraphics[trim={1cm 0 0.5cm 0},width=0.9\columnwidth]{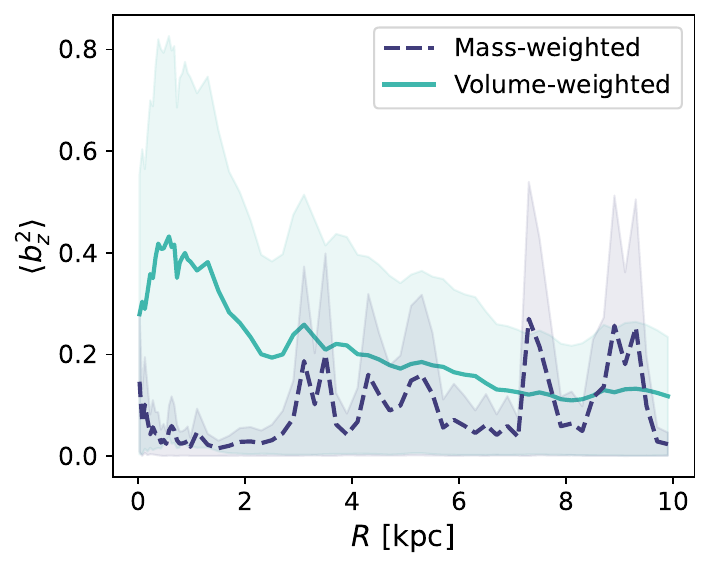}
    \caption{Average value of $b_z^2$ as a function of galactic radius, for two different weightings. Volume-weighted is shown in teal, mass-weighted in dashed dark blue, and our weighted 68\% confidence interval measure plotted in similarly-colored bands. The average is computed within $z=\pm 1$ kpc, with bin sizes $\Delta R = 0.05$ kpc within $R=1 $ kpc, and $\Delta R = 0.2$ kpc outside of $R=1$ kpc. The volume-weighted average falls in general from the center outwards, while the mass-weighted average shows a lot of fluctuations.}
    \label{fig:bzavg}
\end{figure}

Figure~\ref{fig:bzavg} shows the average $b_z^2$ value in radial bins from the center of the galaxy. We calculate the average again within $z=\pm 1$ kpc, with $\Delta R = 0.05$ and $\Delta R=0.2$ kpc within and outside of $R=1$ kpc respectively. The dashed dark blue line is the mass-weighted average, which shows discontinuous spikes in $\langle b_z^2\rangle$. The value of $b_z^2$ remains comparatively low, even at the galactic center. This is consistent with the cold dense ISM for which the magnetic field is strongly toroidal and aligned with the disk. The volume-averaged value, shown in teal, remains close to $0.2$ for all $R$, and approaches even $\langle b_z^2\rangle=0.5$ at $1$ kpc. This is consistent with the warm and hot gas, in which the magnetic field becomes more tangled and random, leading to a more isotropic tilt angle distribution. The weighted confidence interval we calculate is shown in same-colored bands; the bands mimic the average trends.

\begin{figure*}
    \centering
\includegraphics[width=2\columnwidth]{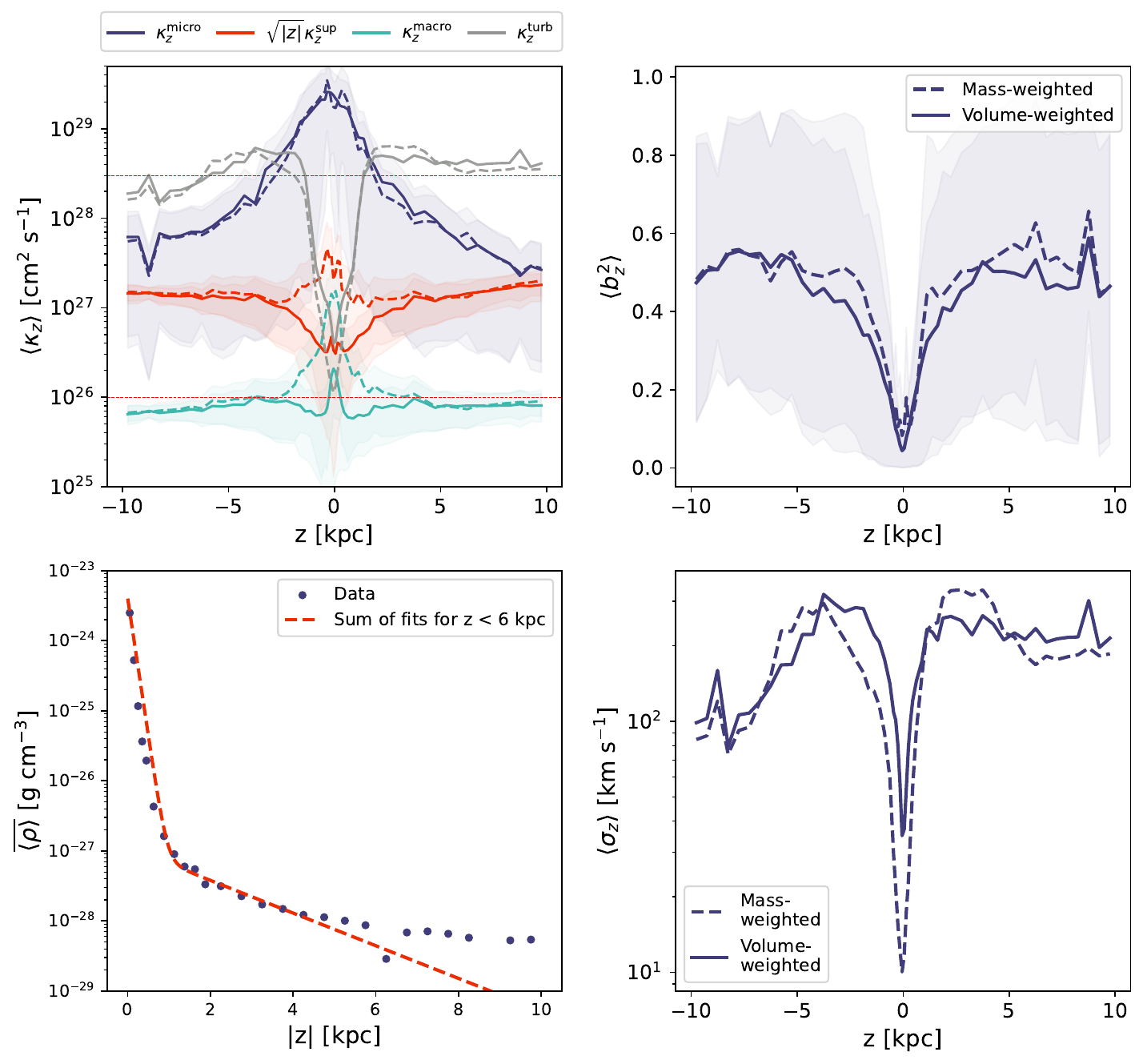}
    \caption{The top left panel shows the diffusion coefficient in the $z$-direction averaged in disk height $z$. From bottom to top we have the macroscopic model in teal, superdiffusive model in red (calculated with the bin's $z$-coordinate as the length scale), and microscopic model in dark blue. In light grey we show the turbulent diffusion coefficient of the gas. The two critical diffusion coefficients we define in Section \ref{sec:critical} are shown as the two horizontal red dashed lines. In the top right panel we show the average $b_z^2$ as a function of $z$.  In the bottom right panel is the average 1-dimensional velocity dispersion in the $z$-direction, which reaches a few hundred km s$^{-1}$ in the galactic corona and drops as low as ${10}$ km s$^{-1}$ near the midplane. For these three panels, line styles correspond to solid lines for volume-weighted averaging, and dashed lines for mass-weighted. Confidence intervals are shown for each weighted distribution. The bottom left panel shows the vertical gas density profile of the disk and halo in dark blue scatterpoints, found by averaging the density profile for $\pm z$. We fit this data using two exponential functions, the sum of which is shown as a dashed red line.} 
    \label{fig:avgz}
\end{figure*}

In the top left panel of Figure~\ref{fig:avgz}, we have plotted our different $z$-direction diffusion coefficients as a function of $z$. From bottom to top we plot the macroscopic model in teal, superdiffusive model in red, and microscopic model in dark blue. Mass-weighted is shown with dashed lines and volume-weighted with solid lines. 

For the superdiffusion model we plot $\sqrt{|z|}\langle k_z^{\rm{sup}}\rangle$, where we have averaged the superdiffusive coefficient $\kappa_z^{\rm sup}$ and then used the cell's $z$-coordinate as the length scale to create an effective diffusion coefficient. This boosts the average diffusion coefficient above $\kappa_z^{\rm{macro}}$ by an order of magnitude, to around $10^{27}~\rm cm^2 s^{-1}$. Note that as $z\rightarrow 0$, we should have $\sqrt{|z|} \kappa^{\rm{sup}} \rightarrow 0$, but this effect is washed out by the averaging process.

While the macroscopic and superdiffusive models have a clear exponential profile within the disk, they level out to approximately constant values in the halo. The microscopic model has a much larger scale height, also significantly larger than the scale height of the disk gas. It falls from $3\times 10^{29}~\rm cm^2 s^{-1}$ in the midplane to $10^{28}~\rm cm^2 s^{-1}$ at $z\simeq 5$ kpc. 

In light grey on this panel we have also plotted the turbulent diffusion coefficient in the $z$-direction, $\kappa_z^{\rm{turb}}$. Recall from Section~\ref{sec:critical} that the turbulent diffusion coefficient is defined as $\kappa_{\rm{turb}}=L\sigma /3$ for some length scale $L$ and velocity dispersion $\sigma$. We calculate the diffusion coefficient associated with resolved (rather than sub-grid) turbulence in the $z$-direction in order to see how entrained we expect CRs to be in the gas. First, to calculate $\sigma_z$ we find the weighted standard deviation in $v_z$ (the $z$-velocity of the gas) in a given bin, which we show in the bottom right panel of Figure~\ref{fig:avgz}.  $\sigma_z$ ranges from $\sim 10~\rm{km s^{-1}}$ in the disk to $\sim 100 ~\rm{km s^{-1}}$ in the corona, as to be expected in these two regimes.

For the length scale $L$ we use the gas scale height $H_g$ as a function of $z$. To calculate $H_g(z)$, we first calculate the gas density $\rho$ in $z$ bins (the bin sizes vary with $\Delta z=0.1$ for $|z|<0.5~\rm{kpc}$,  $\Delta z=0.25$ for $|z|\in [0.5,2]~\rm{kpc}$ and $\Delta z=0.5$ for $|z|> 2~\rm{kpc}$   out to $z=\pm 10~\rm{kpc}$) by summing the mass in a given $z$-bin and and within $R=10~\rm{kpc}$ and dividing by the volume $V=\pi R^2\Delta z$.  We average the $\rho(z)$ distribution for positive- and negative-$z$ to obtain the trend with $|z|$ shown in the bottom left panel of Figure~\ref{fig:avgz}.

The vertical density profile $\rho(z)$ is approximately a double exponential associated with the disk and the halo respectively. We fit an exponential function of the form $\rho(z) = \rho_0 e^{-z/H}$ separately for $z<1.5~\rm{kpc}$ and for $z\in [1.5,6]~\rm{kpc}$ using \texttt{scipy}, and then sum the two curves together - this is shown as the red dashed line shown in the bottom left panel. We choose to sacrifice fitting data beyond $z=6~\rm{kpc}$ in favour of a better transition at the knee. We then recall the definition for the scale height: $$\frac{1}{H_g} = \frac{1}{\rho}\frac{d\rho}{dz}$$ We simulate a dense grid in $z$ and use our $\rho(z)$ fit to calculate $H_g(z)$ using a discretized derivative $\frac{d\rho}{dz}\simeq \frac{\Delta \rho}{\Delta z}$ and interpolate this calculation at our $z$-bin coordinates. This results in a scale height that is relatively constant in each separate regime with a steep transition between the two.

Finally, we calculate $\kappa_z^{\rm{turb}}$ using $\kappa_z^{\rm{turb}}=H_g(z)\sigma_z/3$. The mass- and volume-weighted averages are shown as dashed and solid grey lines respectively. The two averages are about the same except for near the midplane, where the mass-weighted average drops to lower values than the volume-average. Both approach $10^{26} ~\rm{cm^2 s^{-1}}$ in the midplane, and asymptote to a constant value approaching $3-6\times 10^{28}~\rm{cm^2s^{-1}}$ outside the disk.

Two of our critical diffusion coefficients $\kappa_{\rm turb, disk}=10^{26}~\rm cm^2 s^{-1}$ and $\kappa_{\rm turb, corona}=3\times 10^{28}~\rm cm^2 s^{-1}$, introduced in Section~\ref{sec:critical}, are shown with red dashes horizontal lines. Regions where the diffusion coefficient is below each line suggests that CRs may be trapped within their corresponding regions (disk or corona). 

On the top right panel is plotted the average $b_z^2$ as a function of height, which drops from around 0.5 to below 0.1 as $z$ approaches the midplane. This makes sense since the magnetic field primarily lies in the plane of the galaxy; we expect $b_z$ to be small compared to the other components of the magnetic field vector. Interestingly, $b_z^2$, the scale height $H_g(z)$, and $\sigma_z$ all follow the same trend. They increase steeply from the disk out to the corona, and then becoming approximately constant. This can be attributed to the fact that as we transition out of the disk into the CGM, the gas becomes hotter and less dense, with higher degrees of turbulence and thus a more tangled magnetic field.

\subsubsection{Dependence on gas phase}\label{subsec:phase}

In Section~\ref{sec:ingredients} we defined five gas phases based on temperature and ionization fraction. As discussed in the introduction, the WIM/WHIM/HIM make up the highest volume filling fraction in the galaxy, whereas most of the mass lies in the CNM/WNM. CR transport is dependent on the local plasma conditions and therefore sensitive to ISM phase. This is neglected in most simulations that apply a single diffusion coefficient to the entire system. Furthermore, as discussed in \cite{armillotta21} and \cite{ostriker22}, diffusion and streaming dominate the transport in regions with cold and dense gas. Conversely, advection by thermal gas is the most important mechanism for CR transport in regions of high-velocity, hot ISM. They also found that the scattering rate is dependent on the ionization state of the gas. It is much lower in neutral gas than in ionized gas. Therefore CRs will tend to be confined to regions of CNM/WNM with low ionization fraction, because it is difficult to escape into the surrounding regions of ionized, hot diffuse gas due to the high scattering rate the CRs would encounter.  

\begin{figure*}
    \centering
    \includegraphics[width=1.5\columnwidth]{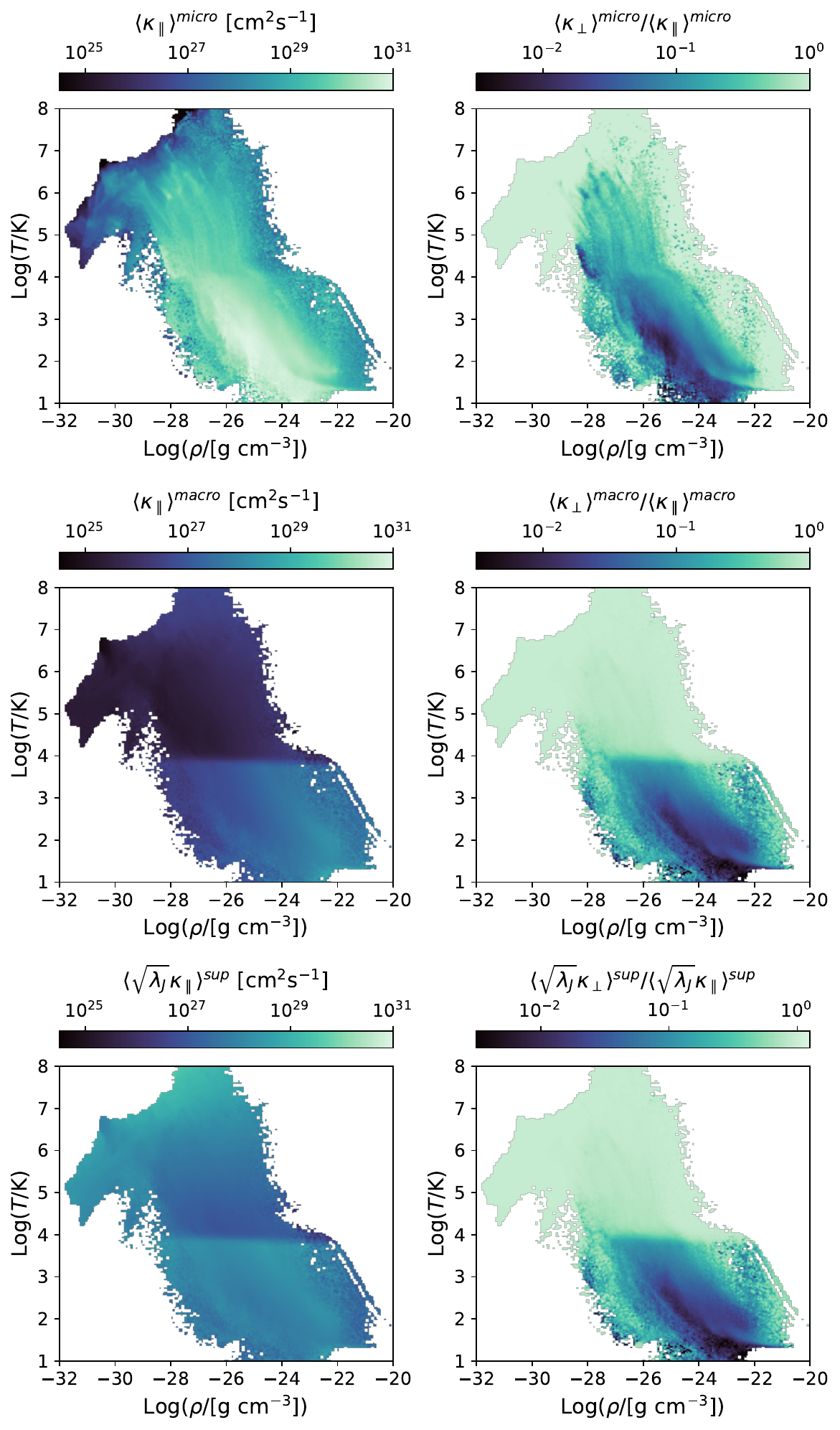}
    \caption{2D histograms of density and temperature from our \texttt{RAMSES} simulation. In the left column each bin's histogram value corresponds to the average diffusion coefficient parallel to the magnetic field. In the right column, the histogram's values are the ratio of the average values of the diffusion coefficients, $\langle \kappa_\perp\rangle/\langle\kappa_\parallel\rangle$. The rows show the microscopic, macroscopic, and superdiffusive model respectively from top to bottom, where the latter uses the thermal Jean's length $\lambda_J$. In general, as expected diffusion is primarily isotropic at high temperatures and anisotropic at low temperatures for the macroscopic and superdiffusive models. The microscopic model shows more variation.}
    \label{fig:phases}
\end{figure*}

In Figure~\ref{fig:phases}, we show 2D histograms of density and temperature (phase space diagrams) from our \texttt{RAMSES} simulation. In the left column, the histogram values correspond to the average diffusion coefficient parallel to the magnetic field. The right column's values correspond to the ratio of the average values of the perpendicular and parallel diffusion coefficient. The three rows show the microscopic, macroscopic, and superdiffusion models from top to bottom respectively. We use the thermal Jeans length $\lambda_J$ as our length scale for the superdiffusive model. All models have primarily isotropic diffusion for high temperature gas. In fact, we see almost completely isotropic diffusion above $T\sim 10^4~\rm K$ for the macroscopic/superdiffusive models. This is to be expected, since the ionization fraction in the simulation is at or above 1 for all temperatures above $\sim 20,000$ K. At high $\chi$ there is significant scattering and CRs have their diffusion directions isotropized between the parallel and perpendicular directions. Below this limit, the macroscopic model has a $\kappa$ ratio that becomes increasingly anisotropic for decreasing temperature, although this is mediated slightly at very high densities.

The microscopic model shows more variation in phase space. Our formulation for this model (see Eqs.~\ref{eq:microiso}, \ref{eq:micropar}, and \ref{eq:microperp}) does not explicitly depend on ionization fraction like the macroscopic model does. Instead, it is a sole function of Alfvén Mach number $\mathcal{M}_{A}$. We can compare the phase-space diagram to the one for $\mathcal{M}_{A}$ in Figure~\ref{fig:phases} and see that the trends indeed mirror each other. We can also compare to the phase-space diagram for $b_z^2$ - super-Alfvénic regions will by definition have gas turbulence that dominates over the magnetic field, leading to more tangled field lines and therefore a propensity for a non-zero magnetic field tilt angle.

\cite{armillotta21} and \cite{ostriker22} simulate CR transport in a post-processing scheme to the multi-phase ISM which includes advection, streaming and diffusion. They find as we do that the value of the diffusion coefficient and the degree of CR trapping vary with gas phase. In \cite{armillotta21}, they find different diffusion coefficients depending on the gas density regime, with scattering rates spanning more than four orders of magnitude. For example, in low density regions ($n_H < 10^{-2}~\rm cm^{-2}$), $\kappa_\parallel\lesssim 10^{28}~ \rm cm^2 s^{-1}$. At high densities where the ionization fraction is low ($n_H > 10^{-2} ~\rm cm^{-3}$), $\kappa_\parallel \gg 10^{29}~ \rm cm^2 s^{-1}$. These also correspond to regimes where different damping mechanisms dominate (nonlinear Landau damping in low densities, ion-neutral damping at high densities). We see a similar trend in the microscopic and macroscopic models wherein the diffusion coefficient increases  with density.

\section{Discussion}
\label{sec:cal}

After carefully analyzing the flow variables that determine the value of the local diffusion coefficients in our three models, we are now in a position to compute the main observables associated with CR transport, namely the gamma ray emission of the galaxy $L_\gamma$ and the midplane CR energy density $U_0$. These properties have been observed to depend on the gas density and star formation activity of the galaxy. For example, \cite{crocker2021a} found that dense starburst galaxies have high degrees of hadronic losses, and results in high gamma-ray output compared to systems with lower star formation rate (SFR) surface densities. They also found that the CR pressure is much lower than gas pressure in starbursts, compared to Milky Way-like galaxies where they are in approximate equipartition. On the opposite side of the galaxy population, in smaller and quieter galaxies, they identify a regime of star formation and gas surface densities in which CRs may drive galactic winds. 

Indeed, a CR's fate is therefore either to be lost to a collision and contribute to the gamma ray emission, or escape out of the galaxy. The ability of CRs to escape the disk depends on the magnitude of the diffusion coefficient perpendicular to the disk midplane, $\kappa_z$, as computed in the previous section. This depends on the ratio of the perpendicular and parallel diffusion coefficients. CRs with higher parallel diffusion coefficients and smaller $\kappa_\perp$ will remain trapped in the plane of the galaxy, especially in regions where the magnetic field is aligned with the disk (small tilt angle), allowing their energy density to build up and the subsequent release of gamma rays. 

In this section we estimate the gamma ray luminosity of GeV CRs that our models predict based on our post-processed \texttt{RAMSES} snapshot. We follow similar analytic calculations to \citet{Krum2020}, hereafter K20, to calculate the so-called calorimetric fraction of our galaxy. The calorimetric fraction is defined here as the ratio of the gamma ray luminosity to the maximum CR energy injected via supernovae. If the calorimetric fraction is one, it means all the available CR energy has been transformed into gamma ray emission.

Following K20 (their Eq.~22), we assume a steady-state relation between the divergence of the vertical CR energy flux and the local CR losses due to proton-proton ($pp$) collisions:
\begin{equation}
    \frac{d}{dz}\Big{(}-\kappa_z\frac{d}{dz}U\Big{)} = -\frac{U}{\tau_{\rm{loss}}}\label{eq:krum22}
\end{equation}
where $\kappa_z$ is the CR diffusion coefficient out of the plane of the galaxy as derived in Section~\ref{sec:kappaz}, $U$ is the CR energy density, and $\tau_{\rm{loss}}$ is the average time a CR travels before being absorbed into the surrounding gas (K20, Eq.23):
\begin{equation}
\tau_{\rm{loss}} = \frac{\mu_pm_H}{\rho\sigma_{\rm{pp}}\eta_{\rm{pp}}c}
\label{eq:tloss}
\end{equation}
where we use the same values for $\mu_p$, $\sigma_{\rm{pp}}$, $\eta_{\rm{pp}}$, and $m_{\rm H}$ as in K20.

As explained before, the gas density in our galaxy follows an approximate exponential profile of the form $\rho=\rho_0\exp{(-|z|/H_g)}$, where $H_g$ is the gas scale height and $\rho_0$ is the mid-plane gas density. For our galaxy, $\rho_0 \simeq 3\times 10^{-24}~\rm{g~ cm^{-3}}$. Putting this in Equation~\ref{eq:tloss} gives
\begin{equation}
    \tau_{\rm{loss}} =  \tau_0 e^{|z|/H_g}
\end{equation}
\noindent where $\tau_{0} \simeq 30~\rm{Myr}$ is the loss time $\tau_0$ at the mid-plane. Equation~\ref{eq:krum22} can finally be rewritten as
\begin{equation}
    \frac{d}{dz}\Big{(}-\kappa_z\frac{d}{dz}U\Big{)} = -\frac{U}{\tau_{0}} e^{-|z|/H_g}
    \label{eq:krum22new}
\end{equation}
\noindent and will be solved for each model separately. If we can approximate $\kappa_z$ as constant, we can pull $\kappa_z$ out of the derivative and divide it over to the righthand side to obtain
\begin{equation}\label{eq:epsilon}
    \frac{d^2U}{dz^2} = \frac{U}{\kappa_z\tau_0}e^{-|z|/H_g}
    =\epsilon \frac{U}{H_g^2}e^{-|z|/H_g}
\end{equation}
where we have defined $\epsilon=\frac{H_g^2}{\kappa_z\tau_0}$. This dimensionless parameter can be used as an indicator of whether or not it is reasonable to neglect hadronic losses for a given model.

We will see in what follows that we can indeed approximate $\kappa_z$ as constant for our models. We have verified this for all three models by also solving Equation~\ref{eq:krum22new} numerically, interpolating our volume-averaged distributions of $\kappa_z$, and find that our answers for the midplane CR energy density $U_0$ are in reasonable agreement.

\subsection{Microscopic diffusion model\label{sec:calmic}}
In our microscopic model, we saw that the average $\kappa_z$ has a scale height $H_{z}\gg H_g$ and thus falls much slower with $z$ than the gas density. Because of this, we can approximate $\kappa_z$ as constant with respect to $z$, since its change from the midplane to within a few $H_g$ of the midplane (our region of interest) will be minor. 

Furthermore, $\kappa_z$ has a $z$-dependence that decreases with increasing $z$, in contrast to the positive exponential with $z$ in K20, so we are unable to use the same analytic evaluation and boundary conditions as they do and instead follow our formulation in Equation~\ref{eq:epsilon}.

For $\kappa_z \simeq 10^{29}~\rm{cm^2 s^{-1}}$, $\tau_0\simeq 30 ~\rm{Myr}$, and $H_g\simeq 100~\rm{pc}$, we have that $\epsilon$ is sufficiently small to neglect hadronic losses and obtain simply:
\begin{equation}
\frac{d^2 U}{dz^2} =0 
\end{equation}
\noindent The CR flux out of the plane in the galaxy (the $z$ direction) is therefore approximately constant and equal to the CRs injected by SNe in the midplane: 
\begin{equation}
    -\kappa_z\frac{\partial U}{\partial z} = -\kappa_z\frac{\partial U}{\partial z}\Bigg{|}_{z=0} = \frac{1}{2}\dot{\Sigma}_*\mathcal{E}_{\rm{SN}}.
\end{equation}
\noindent Here $\dot{\Sigma}_*$ is the SFR per unit area (SFR surface density), and $\mathcal{E}_{\rm{SN}}$ is the specific CR energy injected per supernova, which is $10^{50}$ ergs per $100 ~M_\odot$ in stars formed in the galaxy. The factor of $1/2$ is because half of the energy is released in the $+z$ direction, and half in the $-z$ (see Eq.~26 in K20). 

Solving for $\partial U/\partial z$ and integrating (noting that all the right hand size values are constant with respect to $z$) yields a very simple formula for the CR energy density as a function of height:
\begin{equation}
    U(z) = U_0-\frac{\dot{\Sigma}_*\mathcal{E}_{\rm{SN}}}{2\kappa_z }z \label{eq:Uz}
\end{equation}
\noindent where $U_0$ is the CR energy density at the midplane. Because the diffusion coefficient decreases with $z$ and $U(z)$ cannot become negative, there is a boundary condition where the CR energy density will fall to zero at some height $z_0$, and so we can write Equation~\ref{eq:Uz} as
\begin{equation}
    U(z) = \frac{\dot{\Sigma}_*\mathcal{E}_{\rm{SN}}}{2\kappa_z}(z_0-z). \label{eq:UzBC}
\end{equation}
\noindent We can interpret $z_0$ as a CR scale height with \begin{equation}
    z_0=H_{\rm CR}=\frac{2\kappa_z U_0}{\dot{\Sigma}_* \mathcal{E}_{\rm{SN}}}
\end{equation} 
\noindent that characterizes the magnitude of the CR energy density fall-off with $z$. Let $z_0$ be a multiple of the gas scale height $z_0=\eta H_g$, where $\eta$ is a positive real number. Using this parametrization, the midplane CR energy density is given by
\begin{align}
    U_0 &= \frac{\dot{\Sigma}_*\mathcal{E}_{\rm{SN}}H_g}{2\kappa_z}\eta \label{eq:U0}
\end{align}

As described in Section~\ref{sec:methods}, the disk of our galaxy is defined by
a scale length $R_d \simeq 3.4~\rm{kpc}$ \citep[see][]{Girma}. We compute the SFR surface density $\dot{\Sigma}_*$ by assuming it is constant within $R_d$. We integrate all the mass in stars younger than 100 Myr across the whole galaxy to obtain the total SFR $\dot{M}_*\simeq 0.85~M_\odot \rm{yr^{-1}}$. We divide by the disk surface area $\pi R_d^2$, obtaining $\dot{\Sigma}_*\simeq 0.023 ~M_\odot~\rm{yr}^{-1}~\rm{kpc}^{-2}$.  We take our constant $\kappa_z$ to be that at the midplane, giving $\kappa_z\simeq 2.6\times 10^{29}~\rm{cm^2s^{-1}}.$ Plugging these into Equation~\ref{eq:U0} gives
\begin{equation}
    U_0 \simeq 4.6 \times 10^{-14} ~\eta~\rm{erg}~\rm{cm}^{-3}
\end{equation}
\noindent Now we must determine the value of $\eta$. In Section~\ref{sec:critical}, we have identified two critical turbulent diffusion coefficients: the first one on the scale of the disk (dominated in volume by the WNM and the WIM) and the second one associated with the galactic corona (or the CGM). We can estimate that the height at which the average $\kappa_z$ value in the galaxy falls below any given critical turbulent diffusion coefficient gives the value for $z_0$ that corresponds to CR confinement, and subsequently $\eta=z_0/H_g$. 

In Figure~\ref{fig:avgz} we showed mass- and volume-weighted averaged diffusion coefficients as a function of z. The horizontal lines show the two critical diffusion coefficients. For the microscopic model, $\langle \kappa_z\rangle$ is always above our defined critical value $\kappa_{\rm{turb,disk}}$, meaning CRs will easily escape the disk in which they are injected from SNe. Further emphasizing this point, $\kappa_z$ is above the turbulent diffusion coefficient (shown in light grey) for the inner few kpc, so we expect that the CRs will escape the disk but be confined within the corona. $\langle \kappa_z\rangle$ falls below $\kappa_{\rm{turb,corona}}$ below $z_0\sim -3.4 ~\rm{kpc}$ and above $z_0\sim 2.1~\rm{kpc}$, which we average to get $\eta\sim 27.5$. 

The midplane CR energy density for the microscopic model is thus approximately $U_0\simeq 1.3\times 10^{-12} ~\rm{erg~cm}^{-3}$ or $U_0\simeq 0.78~\rm{eV~cm}^{-3}$. As discussed in the introduction, observations of the Milky Way show that the disk energy density in CRs is in approximate equipartition with those of turbulence, magnetic and thermal energy, at values of $U\simeq 1~\rm{eV ~cm}^{-3}$. Our calculation therefore is in good agreement with observational estimates. We can also speculate that the mid-plane energy density might be low enough that CRs will not radically change the dynamics of our simulated galaxy.

The CR escape fraction is typically defined as $f_{\rm{esc}} = \lim_{z\rightarrow \infty}\frac{\Phi(z)}{\Phi(0)}$. For our model the energy density $U(z)$ falls to $0$ at $\pm z_0$ so the escape fraction becomes $f_{\rm{esc}}=\frac{\Phi(z_0)}{\Phi(0)}$ and is therefore about $1$ and the calorimetric fraction $f_{\rm{cal}}=1-f_{\rm{esc}}\simeq 0$. We can verify this by estimating the gamma ray luminosity we would expect, by integrating the source function $\mathcal{S}$:
\begin{equation}
    L_\gamma = \pi R_d^2\int_{-\infty}^{+\infty} \mathcal{S} dz
\end{equation}
where we use
\begin{equation}
    \mathcal{S}= \frac{U}{\tau_{\rm{loss}}} = \frac{U}{\tau_{0}}e^{-|z|/H_g} 
\end{equation}
After some algebra, we get:
\begin{align*}
     L_\gamma &= \frac{\dot{\Sigma}_*\mathcal{E}_{\rm{SN}}}{2\kappa_z\tau_0}\int_{-z_0}^{+z_0} (z_0-|z|) e^{-|z|/H_g}dz \\
     &= \pi R_d^2\frac{\dot{\Sigma}_*\mathcal{E}_{\rm{SN}}}{\kappa_z\tau_0}H_g^2\Big{[}e^{-z_0/H_g}-1+\frac{z_0}{H_g}\Big{]} \\
     &\simeq 6.5\times 10^4 L_\odot,
\end{align*}
\noindent which is around 1\% compared to the total luminosity of CRs injected into the galaxy $L_{\gamma,\rm{max}} = 2\Phi(z=0)\pi R_d^2 = \dot{\Sigma}_*\mathcal{E}_{\rm{SN}}\pi R_d^2 \simeq 7\times 10^6 L_\odot$. We can directly compute the calorimetric fraction using our simple model as 
\begin{equation}
f_{\rm cal} = \frac{L_{\gamma}}{L_{\gamma,{\rm max}}} \simeq \epsilon \eta = \frac{H_g H_{\rm CR}}{\kappa_z \tau_0}
\label{eq:fcalnew}
\end{equation}
\noindent It is indeed very low for the microscopic model for which $\epsilon \simeq 10^{-3}$ and $\eta \simeq 10$, so that $f_{\rm cal} \simeq 10^{-2} $.

\subsection{Macroscopic diffusion model}\label{sec:calmac}

For the macroscopic model, the diffusion coefficient falls off quickly from the peak with a scale height comparable to that of the disk’s gas density. As we can see in Figure~\ref{fig:avgz}, $\kappa_z^{\rm{macro}}$ falls off by about an order of magnitude for the mass-weighted average, but only a factor of two for the volume-weighted average and then is approximately constant outside of the inner kiloparsec. Therefore we approximate $\kappa_z$ as about constant and equal to the volume-averaged value outside of the disk, which is $\kappa_z\simeq 7.8\times 10^{25}~\rm{cm^2 s^{-1}}$.

Also notice that in this case $\kappa_z$ is below $\kappa_{\rm{turb,disk}}$ everywhere for the volume-weighted diffusion coefficient, and is far less than the $\kappa_z^{\rm{turb}}$ curve except very close to the midplane. The CRs will thus likely remain trapped in the disk. Their energy density will accumulate, which will probably have a strong feedback effect on the galaxy. We expect that this might cause the disk to puff up and thicken for example.  For the same reason, we cannot neglect radiative losses and assume that the RHS of Equation~\ref{eq:epsilon} is zero like we did for the microscopic model.

To calculate the CR energy density we solve Equation~\ref{eq:epsilon} numerically. For our values of $H_g \simeq 100$ pc, $\tau_0 \simeq 30$ Myr and $\kappa_z \simeq 7.8\times 10^{25} ~\rm{cm^2 s^{-1}}$, we obtain $\epsilon \simeq 1.2$, reinforcing that hadronic losses cannot be neglected. We then enforce the same boundary conditions as before: Firstly, we choose the flux at the midplane to be equal to $\Phi_0=\frac{1}{2}\dot{\Sigma}_*\mathcal{E}_{\rm{SN}}$. Secondly, we enforce that $U(z_0)=0$ at some $z_0=\eta H_g$. Because the CRs are likely to remain trapped in the disk, we choose $\eta = 1$. These two boundary conditions result in a midplane value of the CR energy density of $U_0\sim 1.2\times 10^{-10}~\rm{erg ~cm^{-3}}$ or $\sim 73 ~\rm{eV cm^{-3}}$. This is much higher than traditional observational estimates. Furthermore, the escape fraction $f_{\rm{esc}}=\Phi(z_0)/\Phi(0)$ is $\sim 0.7$, giving a calorimetric fraction $f_{\rm{cal}}\sim 0.3$. This would need to be studied further in simulations to include CR backreaction. The luminosity associated with this energy density distribution is obtained by multiplying $f_{\rm{cal}}$ by the CR energy flux at the midplane $L_{\gamma,\rm max}$ calculated in Section~\ref{sec:calmic}:
\begin{equation}
L_\gamma = f_{\rm{cal}}L_{\gamma,\rm max} \simeq 2.2\times 10^6 ~L_\odot
\end{equation}
This value is larger than that of the microscopic model by more than an order of magnitude.

\subsection{Superdiffusion model}\label{sec:calsup}

For the superdiffusion model, we see in Figure~\ref{fig:avgz} that the effective diffusion parameter defined as $\kappa_z = \sqrt{z} \kappa_z^{\rm sup}$ is approximately constant with height far from the disk (i.e. the galactic corona), with a value around $10^{27}~\rm{cm^2~s^{-1}}$. Close to the disk (within $\sim \pm 3$~kpc), the effective diffusion coefficient is well modelled using a positive exponential function for the volume-weighted average (see Fig.~\ref{fig:avgz}), i.e. $\kappa_z\propto e^{\beta z / H_g}$ for some value $\beta$. Interestingly, this is exactly the conditions required to use the same formulation as in K20, but the outer boundary conditions they adopted is not consistent with our framework based on turbulent confinement of CRs within the galactic corona.

To make an estimation of $U(z)$ for this case, we follow a similar approach to the microscopic model (Sect.~\ref{sec:calmic}). The effective diffusion parameter is above our critical value $\kappa_{\rm{turb, disk}}$ across $z$ but remains below the average $\kappa_z^{\rm{turb}}$, so we expect that there is a larger degree of trapping than for the microscopic model, but less than for the macroscopic model. We choose to use the constant value $\kappa_z = 10^{27}~\rm{cm^2 s^{-1}}$. With this value, we obtain $\epsilon\simeq 0.1$. This is larger than the microscopic model where $\epsilon \simeq 0.01$, so potentially not small enough to justify ignoring radiative losses. We will see whether or not neglecting radiative losses is justified by comparing both the numerical and analytical solutions. We choose a value for $\eta$ where the average $\kappa_{\rm{turb}}$ intersects with $\sqrt{|z|}\kappa_z^{\rm{sup}}$, which is at approximately $z_0 = 0.3 ~\rm{kpc}$ or $\eta\simeq 3$.

We find that whether we use the same formulas of Section~\ref{sec:calmic} which neglects radiative losses or a numerical solution like in Section~\ref{sec:calmac} which takes losses into account, we find a midplane energy density of $U_0\simeq 20~\rm{eV cm^{-3}}$ and a corresponding luminosity of $L_\gamma\simeq 1.2\times 10^6~L_\odot$, about $20\%$ of the CR energy flux $L_{\gamma, \rm max}$, or in other words the $f_{\rm{cal}}\simeq 0.2$.  This is higher than that of the microscopic model but less than the macroscopic model, as we expected given that the diffusion coefficient average lies between the other two in Figure~\ref{fig:avgz}. This $U_0$ is more than an order of magnitude higher than the $1~\rm{eV cm^{-3}}$ value observed in the Milky Way. Accordingly, similar to the previous case we would expect that the CR pressure might have dynamical effects on the galaxy and CR backreaction would need to be studied with CR transport simulations. From these results we can also conclude that approximating radiative losses as negligible was a justified assumption for this model.

\subsection{Effects of advection-dominated transport}

In our analytic formulation, we imposed a boundary condition that the CR energy density falls to zero at some height $z_0$. At this height, the turbulent motion of the gas dominates CR transport over diffusion and streaming, so we posited that they will remain confined within this region.

However, in reality the turbulent motions advecting the CRs will effectively create diffusive motion of its own; this gas transport will likely move the CRs beyond the boundary $z_0$. We therefore expect that instead of an approximately linear profile that falls to zero, the $U(z)$ distribution will have wings beyond $z_0$ created by CR energy density building up in regions where the CRs are entrained in the gas, and where radiative losses are much less likely because of the comparatively low gas density.

At distances sufficiently far from the disk, one could model the corona as a sphere with the disk a point at the center, and describe CR transport by a spherical flux. Beyond a certain distance, CRs will propagate away from the corona in a spherically symmetric geometry. We will be able to explore this further in future studies with a full simulation implementation that includes gas advection.

\subsection{Comparison to observations}

We found in Equation~\ref{eq:fcalnew} that the calorimetric fraction $f_{\rm cal}$ is a simple function of galaxy and CR parameters. To generalize it to other galaxies, we note that the midplane hadronic losses timescale $\tau_0$ scales as the inverse of the midplane gas density. If one assumes that turbulence properties in the corona of different galaxies are similar, we can further assume that the CR scale height will also be similar. Under these assumptions, we can rewrite the calorimetric fraction {\it for the microscopic model only} as
\begin{equation}
    f_{\rm cal} \simeq 10^{-2} \frac{\Sigma_g}{10 M_\odot {\rm pc}^{-2}}
\end{equation}
where we have noted that the gas surface density $\Sigma_g \simeq \rho_0H_g$ for midplane density $\rho_0$. Written this way we see that for a given galaxy, the formulations of our study yield a calorimetric fraction that is solely dependent on the gas surface density. The higher the gas surface density, the higher the hadronic loss rate we would expect since a CR would encounter more particles along its path.

We could also write an equivalent relation depending on SFR using the Kennicutt-Schmidt relation with $\dot{\Sigma}_*\propto \Sigma_g^n$ to see that this means higher SFR galaxies would incur higher loss rates. This is indeed the case in starburst galaxies, for which we see higher calorimetric fractions. The higher the vertical diffusion coefficient, the easier it is for CRs to escape out of the disk. This effect is also modulated by the turbulence in the corona of the galaxies. Indeed, the more turbulent the galactic corona is, the more confined the CRs will be, reducing their scale  height $H_{\rm CR}$.

The models in our study (microscopic, macroscopic, and superdiffusive) result in gamma-ray luminosities in the range of $L_\gamma$ from $10^4~L_\odot$ to $10^6~L_\odot$, corresponding to a wide range in calorimetric fractions from 0.01 to as high as $0.7$ for our maximum possible luminosity (if all the CRs energy injected at the midplane was converted to gamma rays), $L_{\gamma, \rm max}\simeq 7\times 10^6L_\odot$. 

We can compare our results to gamma-ray observations of star forming galaxies. Figure 7 of \cite{Kornecki2020}, shows a range of theoretical predictions for how $L_{\gamma}$ will correlate with SFR, for different diffusion prescriptions, as well as a best fit line. Our galaxy has a value for SFR of $\log_{10}(\dot{M})\simeq 0$ and a $\log_{10}(L_{\gamma,\rm max}/[\rm erg ~s^{-1}])\simeq 40.4$, and our models have luminosities ranging from $\log_{10}(L_\gamma/[\rm erg~s^{-1}])\simeq 38$ to $40$. There is a considerable amount of dispersion around the best fit line of their Figure 7 in the SFR region of our interest (which they discuss in their Section 5). Nevertheless we find that our values lie well within the range of expected $L_\gamma$ values, comparable to galaxies in their figure with similar SFRs like the Milky Way, NGC 4945 and the Circinus galaxy. In our model, the dispersion comes from variations in the gas surface density and in the corona's turbulence.

\section{Conclusions}
\label{sec:conclusions}

In this work, we have created two new subgrid models for CR transport and applied them in a post-processing scheme to an isolated, star-forming MHD galactic disk simulated with \texttt{RAMSES} as a proof of concept. In contrast to typical models for CR transport in galaxy simulations, which apply a constant value for the diffusion coefficient to the entire galaxy that can be fine-tuned to produce an accurate gamma ray output, this subgrid model is unique in that it defines the CR diffusion coefficient to be a function of local gas and MHD properties like the Alfvén mach number $\mathcal{M}_A$, ionization fraction $\chi$, turbulent velocity dispersion $\sigma_{\rm{1D}}$, and an assumed turbulence driving scale $\ell_0$. We therefore can simulate the CR transport throughout the galaxy's evolution more accurately, accounting for the ISM's multiphase nature.

When analyzing the radial or vertical trends of various parameters, we employ two averaging techniques. One is a mass-weighted average and the other a volume-weighted average. Depending on the variable, one of these averages is physically-motivated and connected to fundamental conservation laws. The other is not, but highlights behaviour of a certain phase of the ISM. We find that these two averages often result in radial/vertical averages with the same trend (increasing/decreasing) but can differ significantly in magnitude.

We begin by studying the distributions of key physical properties of our galaxy simulation: temperature, gas density, ionization fraction, turbulent velocity dispersion, Alfvén Mach number, and magnetic field strength. We analyze these parameters using spatial maps, phase-space histograms and radial profiles. We also calculate the gas scale height as a function of radius. We find that our star-forming ($\sim 1~\rm M_\odot yr^{-1}$) galaxy is dominated by cold, dense, relatively neutral gas with a scale height of $\sim 100 ~\rm pc$ and high midplane gas surface density of $\sim 600~\rm M_\odot~pc^{-1}$.  It has a relatively weak magnetic field ($B\sim 10^{-6}$ to $10^{-4}~\rm G$). It has a relatively constant ionization fraction that lies between $10^{-3}$ to $10^{-2}$ for a mass-weighted average, or $\sim 1$ for volume-weighted. We also find that the magnetic field, turbulent velocity dispersion Alfvén Mach number, and gas density all decrease with radius.

We then use the relevant plasma parameters to calculate CR diffusion coefficients parallel and perpendicular to the local magnetic field, $\kappa_\parallel$ and $\kappa_\perp$. We do this separately for two approaches to CR transport: the first is our ``microscopic model" that models the diffusion process as the result of a combination of CRs gyrating along magnetic field lines, making a random walk perpendicular to the magnetic field, and undergoing pitch angle scattering through interactions with turbulent MHD fluctuations. The second, our ``macroscopic model", focusses on diffusive transport on large scales: CRs streaming along magnetic field lines at the ionic Alfvén speed effectively diffuse by nature of the magnetic field lines themselves being bent and transported by turbulent advection. In reality, CR transport is a combination of all of these processes: streaming, diffusion, and advection.

It has been found that CR transport may be better described by superdiffusion than classical diffusion. We therefore calculate a third model that is a modification of the macroscopic model, where we calculate superdiffusive coefficients $\kappa_\parallel^{\rm{sup}}$ and $\kappa_\perp^{\rm{sup}}$. We convert these to effective diffusion coefficients for a specified length scale so that we can compare results across models. These length scales include, for different cases, the scale height of the disk $\ell_{\rm{disk}}$, the thermal Jeans length $\lambda_J$, or the $z$-coordinate of a given cell.

We find values for the diffusion coefficient that spans up to six orders of magnitude depending on the gas phase. Our resulting values can be summarized as follows:
\begin{itemize}
\item For the microscopic model, on average $\kappa_\parallel$ increases radially from $\sim 10^{28}$ to $\sim 10^{30}~\rm cm^2 s^{-1}$. It varies widely with gas phase, falling as low as $10^{25}~\rm cm^2 s^{-1}$ for very low density, high-temperature regions and increases at low temperatures/intermediate densities to as high as $10^{31}~\rm cm^2 s^{-1}$. $\kappa_\perp$ is about constant $\simeq 10^{29}~\rm cm^2 s^{-1}$ outside the inner kpc for this model. The ratio $\kappa_\perp/\kappa_\parallel$ increases with temperature, and is $\sim 1$ at very low ($\lesssim 10^{-4}~\rm cm^{-3}$) and very high ($\gtrsim 10 ~\rm cm^{-3}$) densities. 
\item For the macroscopic model, $\kappa_\parallel$ varies less on average in the disk ($\sim 10^{27-28}~\rm cm^2 s^{-1}$) and decreases with radius, conversely to the microscopic model. It decreases to low values $10^{25-27}~\rm cm^2 s^{-1}$ above $T\sim 10^4~\rm K$ where the gas is completely ionized. In this temperature regime, diffusion is also completely isotropic with $\kappa_\perp/\kappa_\parallel\simeq 1$. $\kappa_\perp$ ranges from $10^{26}$ to $10^{28}~\rm cm^2 s^{-1}$ and is also a decreasing function of radius.
\item For both models, $\kappa_\perp/\kappa_\parallel$ falls from $\sim 1$ to $\sim 10^{-1}$ from the center of the disk outwards.
\item For the superdiffusive model, if we use $L=\ell_{\rm{disk}}$ as the length scale to convert to an effective diffusion coefficient, then the trends with radius are identical to that of the macroscopic model. If we use $L=\lambda_J$, $\kappa_\parallel$ becomes relatively constant with radius and $\simeq 10^{28}~\rm cm^2 s^{-1}$. $\kappa_\perp$ also levels out more, sitting at $\sim 10^{27-28}~\rm cm^2 s^{-1}$ across the disk. In this case, the ratio of the two is about isotropic across the entire disk. Interestingly, we find that these effective classical diffusion coefficients calculated from our superdiffusive model are closer to empirical ones than both our classical microscopic and macroscopic models.
\end{itemize}

We also calculate a diffusion coefficient associated with turbulence, $\kappa_{\rm{turb}}$, along with three critical turbulent diffusion coefficients, $\kappa_{\rm turb, cell}$, $\kappa_{\rm turb, disk}$ and $\kappa_{\rm turb, corona}$, corresponding to the scales of the simulation grid cell, molecular clouds and galactic coronae. If the CR's rate of diffusion out of the plane of the galaxy is lower than $\kappa_{\rm{turb}}$ for a given scale length, the motion of the gas will confine the CRs. The gas is being circulated faster than the CRs can escape the region. We find that $\kappa_z^{\rm turb}$, the resolved turbulent diffusion coefficient in the $z$-direction, rises from $\sim 10^{26}~\rm cm^2 s^{-1}$ at the midplane to above $10^{28}~\rm cm^2 s^{-1}$ outside the disk and then is constant.

When examining the CR flux out of the plane of the galaxy, we find that the diffusion coefficient out of the plane of the galaxy (in the $z$-direction) plays a crucial role in how confined CRs remain in the disk. In general, we expect that the magnetic field lies in the plane of the galaxy. We also assume CR transport is sufficiently efficient to homogenize the distribution of CRs in the disk. We can therefore expect that regions with $\kappa_\perp/\kappa_\parallel \ll 1$ correspond to a higher degree of CR trapping. However, the magnetic field direction is not solely in the plane of the galaxy. Therefore to calculate the upward flux of CR energy we derive a diffusion coefficient in the $z$-direction, $\kappa_z$.

We find that $\kappa_z$ depends explicitly on the orientation of the local magnetic field through the square of the $z$-component of the magnetic field unit vector, $b_z^2$. A critical result of this work is thus that the ability of CRs to escape the galactic disk into the halo and beyond, along with whatever dynamical and energetic impacts this imposes on the system, is directly dependent on the galaxy's magnetic field topology. This is another reason why using the MHD galaxy of \texttt{RAMSES}, which defines the 3D magnetic field vector for every cell, allows for self-consistent simulations. The macroscopic and microscopic models have $\kappa_z$ that is relatively constant except for in the disk, with values $\sim 10^{26}~\rm cm^2 s^{-1}$ and $\sim 10^{27}~\rm cm^2 s^{-1}$ respectively. The microscopic model has a larger scale height and falls from  $\gtrsim 10^{29}~\rm cm^2 s^{-1}$ to $\lesssim  10^{28}~\rm cm^2 s^{-1}$ by $z\simeq 10~\rm kpc$.

Lastly, we aim to calculate observables like the gamma ray luminosity, midplane CR energy density, and calorimetric fraction of the galaxy. We separately derive the CR energy density as a function of height $U(z)$ for each model by i) approximating the vertical diffusion coefficient $\kappa_z$ as constant, ii) identifying $\kappa_z$ as either high enough to render radiative losses negligible (which applies to the microscopic and superdiffusive models), or not high enough such that radiative losses must be taken into account, and iii) solving a steady state equation for the evolution of vertical CR flux to obtain $U(z)$. We impose boundary conditions on the midplane CR flux and at a critical height $H_{\rm CR}$ where $U(z)$ falls to 0. This defines the characteristic CR confinement scale. The CRs will remain trapped within this region, moving at high speeds with long lifetimes before they are lost to $pp$ collisions. Ignoring radiative losses yields a simple solution for $U(z)$ that is a linear function of $z$. When including radiative losses, we have to solve the CR flux equation numerically. The numerical solution turns out to be very close to the analytical approximation.

For the microscopic model the estimated resulting calorimetric fraction is very small ($f_{\rm{cal}}\simeq 10^{-2}$), and the midplane energy density $U_0$ is comparable to observations in the Milky Way . The superdiffusive model has an intermediate calorimetric fraction ($f_{\rm{cal}}\simeq 0.2$) with $U_0$ an order of magnitude above Milky Way observations. The macroscopic model has $\kappa_z < \kappa_{\rm turb, disk}$ everywhere and so we expect the CRs to remain in general trapped in the disk, allowing the energy density of CRs to build up. We indeed find the largest $f_{\rm cal}$ for this model, with $f_{\rm{cal}}\simeq 0.3$ and $U_0 \simeq 70~\rm eV ~cm^{-3}$.

For the macroscopic and superdiffusive cases, we expect that there may be feedback processes where the disk heats and thickens/puffs up, serving to increase the gas scale height and potentially induce a dynamical effect on the galaxy. Therefore what we are calculating in essence is a snapshot of the calorimetric behaviour at a given time, before feedback processes create system evolution and potentially an approach to long-term, semi-steady state behaviour. This lack of simulating CR backreaction is one of the main caveats of our study. For example, we also cannot know for certain whether or not our superdiffusive model actually yields the most accurate diffusion coefficients because feedback processes that would occur as the system evolves in a full simulation (gas outflows, disk heating/thickening, gas advection, etc.) are not explored in this paper. Additional caveats include that we assume steady-state behavior for the CR energy flux equation (Eq.~\ref{eq:krum22}), and that \texttt{RAMSES} assumes a fixed UV background without taking into account the local interstellar radiation field which may slightly affect the ionization fraction calculations.

For all models we find a range of gamma ray luminosities, between $10^4 $ to $10^6 ~L_\odot$. These are comparable to gamma ray observations of the Milky Way and other star forming galaxies with SFRs that lie within an order of magnitude of our galaxy's.

\section*{Acknowledgements}

The authors would like to thank Phil Hopkins and Matt Sampson for helpful discussions and feedback.

This material is based upon work supported by the National Science Foundation under Award Number 10014005 and Award Title ``Physical Origin of Globular Clusters in the Context of Galaxy Formation''.

The simulations presented in this article were performed on computational resources managed and supported by Princeton Research Computing, a consortium of groups including the Princeton Institute for Computational Science and Engineering (PICSciE) and the Office of Information Technology’s High Performance Computing Center and Visualization Laboratory at Princeton University.


\section*{Data Availability}

The simulation data used in this article will be shared upon reasonable request to the corresponding author.
 



\bibliographystyle{mnras}
\bibliography{bib}

\bsp	
\label{lastpage}
\end{document}